\documentclass[12pt,preprint]{aastex}

\usepackage{lscape}

\shorttitle{M82 Star Clusters}
\shortauthors{Lim, Hwang, \& Lee}

\begin{document}


\title{
The Star Cluster System in the Nearby Starburst Galaxy M82} 


\author{Sungsoon Lim\altaffilmark{1}, Narae Hwang\altaffilmark{2} and
Myung Gyoon Lee\altaffilmark{1}}

\email{slim@astro.snu.ac.kr,  nhwang@kasi.re.kr, 
mglee@astro.snu.ac.kr}

\altaffiltext{1}{Astronomy Program, Department of Physics and Astronomy, Seoul
 National University, Seoul 151-747, Korea }
\altaffiltext{2}{Korea Astronomy and Space Science Institute, Daejeon 305-348, Korea }





\begin{abstract}

We present a photometric study of star clusters  in the nearby starburst galaxy
M82 based on the \textit{UBVI}, \textit{YJ} and \textit{H} band {\it Hubble Space Telescope} images.
We find 1105 star clusters with $V<23$ mag. 1070 of them are located in the disk region, while 35 star clusters are in the halo region.
The star clusters in the disk are composed of a dominant blue population with a color peak at $(B-V)_0 \approx 0.45$, and a weaker red population. 
The luminosity function of the disk clusters shows a power-law 
distribution with a power law index $\alpha=-2.04\pm0.03$, 
and the scale height of their distribution is $h_z=9\arcsec.64\pm0\arcsec.40$  ($164 \pm 7$ pc), similar to that of the stellar thin disk of M82.
We have derived the ages of $\sim$630 star clusters using the spectral energy distribution fit method by comparing \textit{UBVI(YJ)H} band photometric data with the simple stellar population models.
The age distribution of the disk clusters shows that the most dominant cluster population has ages ranging from 100 Myr to 1 Gyr, with a peak at about 500 Myr. 
This suggests that M82 has undergone a disk-wide star formation about 500 Myr ago, probably through the interaction with M81.
The brightest star clusters in the nuclear region are much brighter than those in other regions, indicating that more massive star clusters are formed in the denser environments.
On the other hand, 
the  colors of the halo clusters are similar to those of globular clusters in the Milky Way Galaxy, and their ages are estimated to be older than 1 Gyr.
These are probably genuine old globular clusters in M82.
\end{abstract}





\section{Introduction}

Nearby starburst galaxies have been of great importance in terms
that they provide a detailed view of the starburst process in galaxies. Especially,
thanks to the high spatial resolution images obtained with the {\it Hubble Space Telescope}
(HST), it became possible to study the properties of the star clusters and to use them for investigating the star formation history in these galaxies.
There are numerous studies of young star clusters in starburst or
star-forming galaxies (e.g. \citealt{Whi99,Lar02,Sch07,Par09,Hwa08,Can09,Hwa10,
Pel10,Whi10,Cha10,Lar11}).
It is found that  young star clusters in these galaxies show a power law luminosity function ($N \propto M_V^\alpha$)
 with similar power index values ($\alpha \sim -2$) \citep{bas10,Hwa08}, and that
 some  of them  are more massive
($10^5M_\odot\lesssim M \lesssim 10^6M_\odot$) than normal young
clusters ($M \lesssim 10^4M_\odot$) so that they are called super star clusters (SSCs).

Existence of an old halo in these galaxies also has been known. 
Several studies investigated old star clusters 
in star-forming or starburst galaxies. 
A significant population of old star clusters is found even in late-type galaxies with high star formation rate
(e.g. \citealt{Cha04}).
\citet{Hwa08} found several tens of old globular cluster 
candidates in an interacting and highly
star-forming spiral galaxy M51, and estimated its globular cluster specific frequency to be as low as
$S_N \sim 0.25$, which is even lower than that of the Milky Way ($S_N \sim 0.5$).
\citet{Whi10} reported the presence of some old clusters in the Antennae galaxy with high star formation activity.

M82 is one of the nearest starburst galaxies, and its star formation rate
is as high as $\sim10$M$_{\odot}$ yr$^{-1}$ \citep{OC78,Bel01}, one of the highest
values in the nearby universe (see \citet{May09} for an extensive review). M82 has been known as Irr II or Amorphous for long. However discovery of a bar and spiral arms in the near-infrared (NIR) images led it to be classified as SBc \citep{tel91, May05}. 
It has a dynamical mass of $\sim10^{10}$ $M_\odot$ \citep{gre12}. 
This galaxy belongs to the M81 Group and it is
embedded in a huge neutral hydrogen cloud connecting all major member galaxies
\citep{Yun94,Chy08}. The observation of group-wide neutral hydrogen structure
provided a strong evidence that the interaction between M81 and M82 might have induced
the starburst phenomenon in M82 \citep{rie80,Yun94,Yun99,For03}.

There have been many studies on the star clusters in M82
since \citet{OC78} reported the existence of several knots and semi-stellar objects in the central region of M82. 
Some of them were  classified as SSCs 
 based on their high luminosity ($-13.2 \leq M_{B} \leq -9.6$) \citep{OC95}.
Later \citet{Mel05} detected 197 SSCs with $17 \lesssim F555W \lesssim 21$ in the nuclear region of M82 using the data taken with $HST$/Wide Field Planetary Camera 2 (WFPC2), and
concluded that most nuclear star clusters in M82 are young (age $<25$ Myr) and compact ($r=5.7\pm1.4$ pc).
\citet{Bar08} reported an enhanced star cluster formation at about $100$ Myr ago in the inner few hundred parsecs of M82 based on the age distribution of about 150 star clusters.

Other than the central or nuclear region, a post-starburst region B has been a target of many studies. 
\citet{deG01} detected 113 star clusters in the region B using the data taken with
$HST$/WFPC2 and Near Infrared  Camera and Multi-Object
Spectrometer (NICMOS) 
and concluded that the last tidal encounter between M81 and M82 occurred at
about 600 Myr ago based on the age distribution of the star clusters. Later \citet{deG03} revisited the same data set and argued that the encounter epoch is about 1 Gyr ago. 
However, \citet{Smi07} found 35 star clusters in the region B using the image data obtained with $HST$/Advanced Camera for Surveys (ACS) and concluded that M81 and M82 might have encountered at about 150 Myr ago based on the ages of the detected star clusters.
This result is supported by \citet{Kon08} 
who showed that the spectroscopic ages of seven star clusters in the region B are in agreement with those reported in \citet{Smi07}.

A study of the star clusters over the entire disk of M82 was given  by \citet{May08} who found 653 star clusters in the disk of M82 using 
the image data obtained with $HST$/ACS. However, they assumed two fixed values for the ages of star clusters, 8 Myr and 100 Myr, instead of deriving their ages, to study cluster mass distributions. 
Therefore, it is still needed to make a comprehensive study on the age distribution of star clusters over the entire disk of M82 to investigate its star cluster formation history. 
Since numerous previous studies on the same issue have led to various conclusions regarding the epochs of the encounter between M81 and M82, a study on the ages of the star clusters in the entire M82 disk, not restricted to some specific regions, will be of great merit.

Another important issue is to find old halo clusters in M82 and to investigate their properties.
To date only a few old globular clusters were reported from a couple of spectroscopic observations of limited number of candidates \citep{Kon09,Sai05} so that little is known about the old population in M82.
Because of its proximity, M82 is an ideal target to search for old halo populations, if any. 
M82  is a very useful galaxy for 
understanding the existence and properties of old stellar populations in starburst galaxies. 

In this study, we present a survey of the star clusters in  M82 using high resolution  optical and near-infrared ($U$ to $H$) images taken with $HST$/ACS Wide Field Channel (WFC) and $HST$/Wide Field Camera 3 (WFC3), covering not only the disk but also some halo area in M82.
The addition of $U$ and $H$ band data improves the capability to deal with the reddening and to reduce the age estimation uncertainty for star clusters.
This paper is organized as follows: The methods of data reduction and analysis are
described in Section 2. \S 3 presents the main results. Implication of
these results are discussed in \S 4, and the primary results are
summarized in \S 5.

We adopt a distance to M82, 3.55 Mpc ($(m-M)_0=27.75\pm0.07$)  determined from
the measurement of the $I$-band magnitude of the tip of the red giant branch (TRGB) magnitude \citep{Lee13}.
The distance to M82 is similar to that to M81, $3.63\pm 0.14$ Mpc ($(m-M)_0= 27.80\pm 0.08$) \citep{dur10,ger11}.
At this distance,
one arcsecond (arcmin) corresponds to 17 pc (1.02 kpc) in physical scale.
The foreground
reddening toward M82 is $E(B-V)=0.159$ and the corresponding extinctions are
$A_U=0.862$, $A_B=0.685$, $A_V=0.526$, $A_I=0.308$,  $A_{YJ}=0.177$, and $A_H=0.091$ \citep{Sch98}. The total
magnitudes, colors, and position angle of M82 are $B_T=9.3\pm0.09$,
$(B^T-V^T)=0.89\pm0.01$, and 65 degrees \citep{deV91}. At the adopted distance
(without the internal extinction correction), the absolute magnitudes are
$M_B ^T=-19.13$ and $M_V ^T =-19.87$.
The inclination angle of M82 is
$i=81.5$ degrees in the \textit{V} band \citep{Lyn63}, $i=73$ and $77\pm3$ degrees in the $H$ and $Ks$ band \citep{Ich95},
and $77\pm3$ degrees in the $JHKs$ images \citep{May05}, respectively.
We adopt $77\pm3$ degrees for analysis in this study.
We use the center position of M82 from NASA/IPAC Extragalactic Database 
based on the Jordell--VLA Astrometric Survey (JVAS) and Cosmic Lens All-Sky Survey (CLASS). 

\section{Data Reduction \& Analysis}

\subsection{Data}

Table \,\ref{tbl-1} lists a summary of the data used in this study. 
We used the image data obtained by the Hubble Heritage Team using the $HST$/ACS
in $F435W$, $F555W$, and $F814W$ filter bands
(called  $B$, $V$, and $I$ hereafter) 
(proposal ID 10776, PI: M. Mountain). In each filter band, six exposures were taken in a 3$\times$2 mosaic
configuration with the accumulated exposure times of 1600, 1360, and 1360
seconds, respectively. The final field coverage is about $12\arcmin \times
8\arcmin$ centered on M82, covering the optical disk of M82. All the necessary
data reduction processes were done by the Hubble Heritage Team, including the
ACS image data reduction with the standard ACS pipeline (CALACS) and the image
combination with MultiDrizzle task. More detailed information on the
observation and data reduction is given in \citet{Mut07}.
The full width at half maximum
(FWHM) of a point source in the $V$-band image is about 0.1 arcsec.

\begin{figure}
 \epsscale{1.0} \plotone{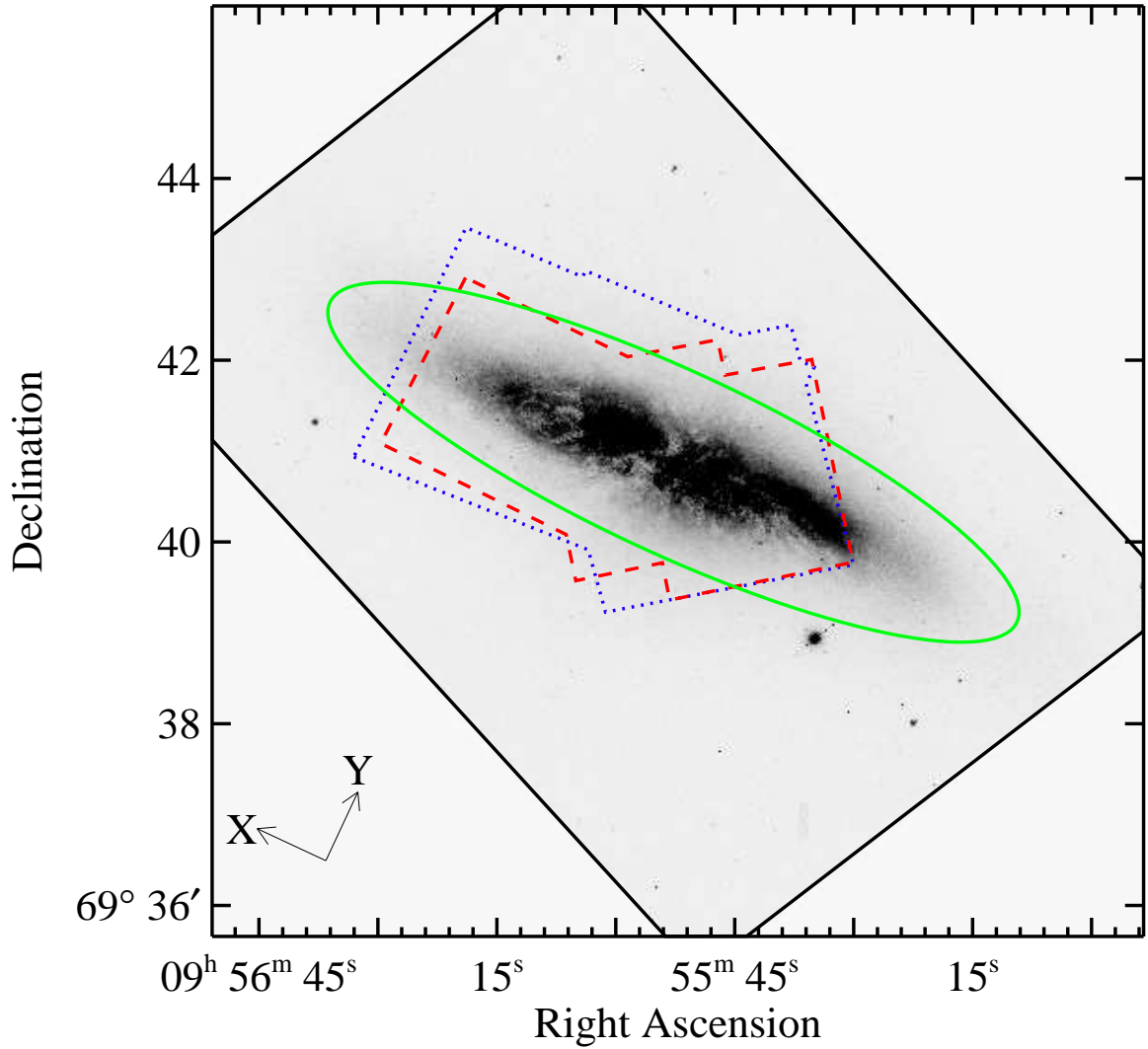} 
 \caption{A grey scale map of the $HST$/ACS $I$-band image of M82.
 Solid, dotted, and dashed lines represent the areas covered by $HST$/ACS ($F435W(B)$, $F555W(V)$ , and $F814W(I)$), $HST$/WFC3 UVIS ($F336W(U)$), and
 NIR ($F110W(YJ)$ and $F160W(H)$) images, respectively. 
 A large ellipse represents the survey area of \citet{May08} with $250\arcsec$ of major axis, $60\arcsec$ of minor axis, and 65 degrees of position angle.
 Arrows at the bottom-left corner represent $X$ and $Y$ 
 directions defined in Section 3.2.
\label{finder}}
\end{figure}

We also used the ultraviolet (UV) and near-infrared (NIR) image data taken with $HST$/WFC3 (Proposal ID
11360, PI: R. O'Connell) that partially covered the M82 disk in $F336W$ ($U$),
$F110W$ (\textit{YJ}), and $F160W$ ($H$) bands.
The full width at half maximum
(FWHM) of a point source in NIR images is about 0.25 arcsec.
All the image data provided from the $HST$ archive are processed through the
`on-the-fly' calibration based on the best reference frames available. We have
combined three exposures for $U$ and two exposures for \textit{YJ} and \textit{H} using SWARP \citep{Ber02} to make a mosaic image in
each filter band. The final mosaic images cover about $5 \arcmin \times 2
\arcmin$. See Figure \,\ref{finder} for the areal coverage of each band.

\subsection{Source Detection and Photometry}

Since M82 is an almost edge-on disk galaxy,
it is very challenging to detect sources in the disk and central region of the galaxy.
We combined the $B$, $V$, and $I$ band images, and applied
a ring median filtering method (with filter radii $r_{in}=15$ and $r_{out}=20$ pixels) with RMEDIAN task in IRAF to the combined image
to generate a map of smoothly varying background brightness. Then we subtracted it
from the original combined image, obtaining a master detection image.
This method improves the detection efficiency for the sources in the disk.
A similar method was adopted by \citet{Hwa08} successfully to detect the star clusters buried in the spiral arms of M51.
We used Source Extractor \citep{Ber96} on the master detection image to detect
sources with a threshold of 3$\sigma$ and a minimum contiguous detected area of
five pixels.

We used PHOT task in IRAF/DAOPHOT package to derive photometry in
$UBVI$, \textit{YJ}, and $H$
band images at the coordinates of detected sources. A single aperture with fixed
size of $r=0.35\arcsec$ was adopted in all bands for the derivation of consistent
colors. The aperture size was determined based on the FWHM of point sources in the
NIR WFC3 image data with FWHM$\approx0.3\arcsec$, which provides the poorest resolution in our data set.
The photometric zero points for $HST$/ACS and WFC3 in Vega magnitude system are
adopted from the STScI websites
(http://www.stsci.edu/hst/acs/analysis/zeropoints/$\#$tablestart, \\ and http://www.stsci.edu/hst/wfc3/phot$\_$zp$\_$lbn, respectively).

The aperture magnitude requires correction for the lost light fraction due to
the finite size of an aperture. The aperture correction for point sources is
determined based on the instrumental characteristics of the $HST$. However,
for extended sources, the aperture correction also depends on their light profile
characteristics and sizes. Therefore, we use an aperture correction method for
extended sources with a circular shape, as adopted by \citet{Hwa08}, based on
the correlation between the slope of light profiles and the calculated value for
the aperture correction. This correlation is obtained by calculating the slope of
the aperture magnitude profiles between $r=5$ pixels ($=0.25\arcsec$)
and $r=7$ pixels ($=0.35\arcsec$) for well-isolated and extended sources with
circular shapes. The aperture correction values are calculated by difference
between aperture magnitudes with $r=7$ pixels ($=0.35\arcsec$) and $r=15$ pixels
($=0.75\arcsec$). There is a good correlation between the slope and the aperture
correction value ($\Delta $Mag$ = 2.94 \times $slope$-0.13$, rms$=0.09$). To avoid over correction, we restrict the maximum correction
value as 1.5 mag that is the largest aperture correction value of well-isolated
sources. 
These aperture correction values are estimated only 
on the $V$-band image to derive the $V$-band total magnitudes of the star clusters.

\subsection{Star Cluster Selection}

Since selecting star clusters in nearby galaxies  is a challenging process even with the high resolution $HST$ images, various methods have been used for the $HST$ images of nearby galaxies.
\citet{Sch07} selected star clusters in M51 based on the source
sizes. They measured the size of the sources using ISHAPE \citep{Lar99}, and
selected star cluster candidates with effective radii $r_{\rm eff}>0.5$ pc.
Then they manually inspected the candidates for final selection.
\citet{May08} selected star clusters in M82 using the parameters provided by
Source Extractor. They checked FWHM, isophotal area,
ellipticity, and magnitude of
the detected sources to select star cluster candidates. They also finalized
the star cluster candidates through visual inspection.
\citet{Hwa08} selected star clusters in M51 using a combination of automatic screening and visual inspection.
They used the FWHM and ellipticity information of the sources to define the
cluster candidates. Star clusters were selected by inspecting radial profiles
and 2D contours of the candidates. Both the automatic method and the visual inspection
method are successful in cluster finding. However, the automated methods have
difficulties in finding the clusters located in the varying background.
It needs additional visual check. Star clusters in M82 are often located in the bright and
crowded background (especially in the disk),
so that visual inspection is an indispensable step to find star
clusters in M82.
Therefore we use a combination of automatic screening and visual inspection
to search for star clusters in M82 in this study.
First we made a list of about 14000 candidates for visual inspection, selecting the sources
with FWHM$>2.5$ pix and $V<23$ mag (after aperture correction).
Then we visually checked the images, radial profiles, and isophote contour maps of the sources.
Finally we selected 1105 star clusters, satisfying the following criteria: (1) round appearances, (2) extended radial profiles, and (3) lack of smooth extended structures.

\begin{figure}
 \epsscale{1.0} \plotone{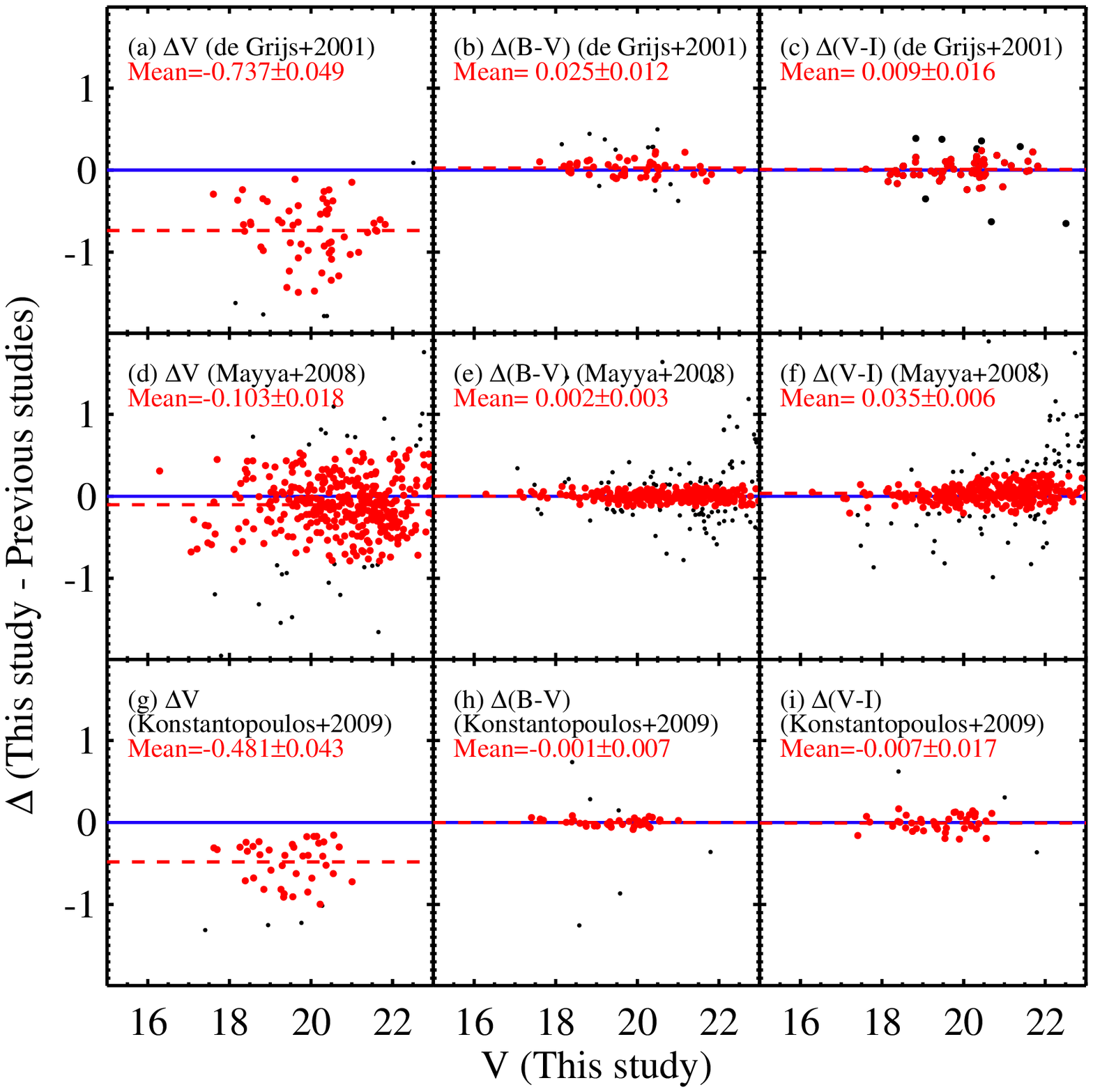} 
 \caption{Comparison of our photometry of M82 star clusters with those in previous studies, \citet{deG01} (upper panels),
 \citet{May08} (middle panels), and \citet{Kon09} (lower panels).
 Left, middle, and right panels represent the differences in $V$, $(B-V)$, and $(V-I)$, respectively. 
  $\Delta$ represents this study minus previous studies. 
 Filled circles  represent the data used for deriving average values (dashed lines) with 2-sigma clipping.
\label{compphot}}
\end{figure}

We compared our photometry of the M82 star clusters with those in previous studies \citep{deG03, May08, Kon09},
as shown in Figure. \ref{compphot}.
The $(B-V)$ colors in these four studies show a very good agreement.
$V$ magnitudes in this study show an approximate agreement with those in \citet{May08},
but are systematically brighter than those in other studies \citep{deG03, Kon09}. 
This is due to the fact that the magnitudes
in this study and \citet{May08} provide total magnitudes for star clusters, while the other studies provide magnitudes for the fixed apertures.
The mean differences derived with 2 sigma clipping are
$\Delta V =-0.737 \pm0.049 $, $\Delta (B-V) = 0.025\pm0.012$,  and $\Delta (V-I) =0009 \pm0.016$ for \citet{deG03} minus this study,
$\Delta V = -0.102\pm0.018 $ , $\Delta (B-V) =0.002 \pm0.003$,  and $\Delta (V-I) = 0.035\pm0.006$  for \citet{May08} minus this study, and
$\Delta V = -0.474\pm 0.043$ , $\Delta (B-V) = -0.001\pm0.007$,  and $\Delta (V-I) = -0.008\pm0.016$  for \citet{Kon09} minus this study.
\citet{May08} adopted the isophotal magnitudes
provided by Source Extractor, which are similar to the total magnitudes.

\subsection{Size Estimation of the Star Clusters}

To measure the size of star clusters, we used ISHAPE that was designed to estimate the size of slightly extended sources in images \citep{Lar99}.
It assumes that the image of a source can be modeled as convolution of a point spread function
(PSF)  and analytic functions for an extended source.
We made a PSF image using DAOPHOT/SEEPSF task in IRAF for selected bright isolated stars. We adopted a MOFFAT15 model for analytic function.
ISHAPE provides FWHMs of the sources,
and we converted them to effective radii by multiplying 1.13,
according to the guide of the ISHAPE manual.

\subsection{Estimation of Age, Extinction, and Mass of the Star Clusters }

We have estimated ages, extinctions, and masses of the star clusters in M82
 by comparing the photometric data with
the model spectral energy distributions (SEDs). Model SEDs are derived from the simple stellar population (SSP) synthesis models by \citet{BC03}. The models span ages from 1 Myr to 15 Gyr for metallicity range of Z$=0.0001-0.05$. 
We assume a Salpeter initial mass function (IMF) of $N(m)dm \propto m^{-2.35}$ \citep{Sal55} with a lower mass limit with 0.1 $M_{\odot}$ and an upper mass limit with 100 $M_{\odot}$. 
We adopted a solar metallicity for the star clusters in the disk based on the abundance of HII regions in M82 \citep{Smi06}.
Note that using a Salpeter IMF could overestimate the cluster mass by a 
factor of 1.70--3.46 although relative mass distribution of star clusters is not changed \citep{deG03}.
We adopted the Salpeter IMF, because most of the previous studies adopted the same IMF and because we wanted to compare our results with them.  We checked the effect of adopting the Chabrier IMF in comparison with the Salpeter IMF for SSP models in the $(U-B)-(B-V)$ color-color diagram. 
The colors of the SSP models with these two IMFs are almost same for log $t>8$, and they change by ~0.1 for log $t<7$. Therefore our results are affected little by adopting different IMFs.

SED fitting based on broad band photometry is known to have some limits, but
is still very useful for estimating ages of star clusters.
\citet{And04} showed that at least four band data are needed to obtain reliable
age estimation with an SED fit method. Especially, $U$ and $B$ bands are important
for age estimation of young star clusters, because those bands cover the Balmer
jump that is a sensitive age indicator. \citet{Mai09} also suggested that
the combination of $U$, optical, and NIR band data is useful to determine the ages of
star clusters when the extinction of each cluster is unknown. M82 is almost
edge-on, and most star clusters in the disk suffer from significant extinction.
We utilize six $UBVI(YJ)H$ photometric bands for age and extinction estimation of the
star clusters in the disk region.
About 870 star cluster candidates in the central region of M82 were covered in six bands.
However, only $BVI$ images are available for about 200 star clusters, which are located mostly in the outer region of M82.

The estimated ages of the clusters are not affected by the cluster size estimation, because the cluster colors were derived from the fixed aperture photometry. It is also noted that the estimated masses of the clusters do not depend on the cluster size estimation, because they were derived from the $V$-band total magnitudes.

\section{Results}

\subsection{A Catalog of the Star Clusters in M82}

In Table\,\ref{tbl-2}  we list a catalog of 1105 star cluster candidates (hereafter star clusters) with $V<23$ mag. 
Brief descriptions on the data columns of the catalog are given as follows:
Column 1 lists an identification number of a star cluster;
Columns 2 and 3 list J2000.0 right ascension (RA) and declination (Dec) in degrees;
Columns 4 lists aperture corrected $V$-band magnitudes and corresponding errors;
Columns from 5 to 9 list ($B-V$), ($V-I$), ($U-B$), ($I-YJ$), and ($YJ-H$) colors and corresponding errors for an aperture radius $r=0\arcsec.35$;
and Column 10 and 11 list effective radii and ages, respectively.

\subsection{Spatial Distribution of the Star Clusters}

\begin{figure}
 \epsscale{1.0} \plotone{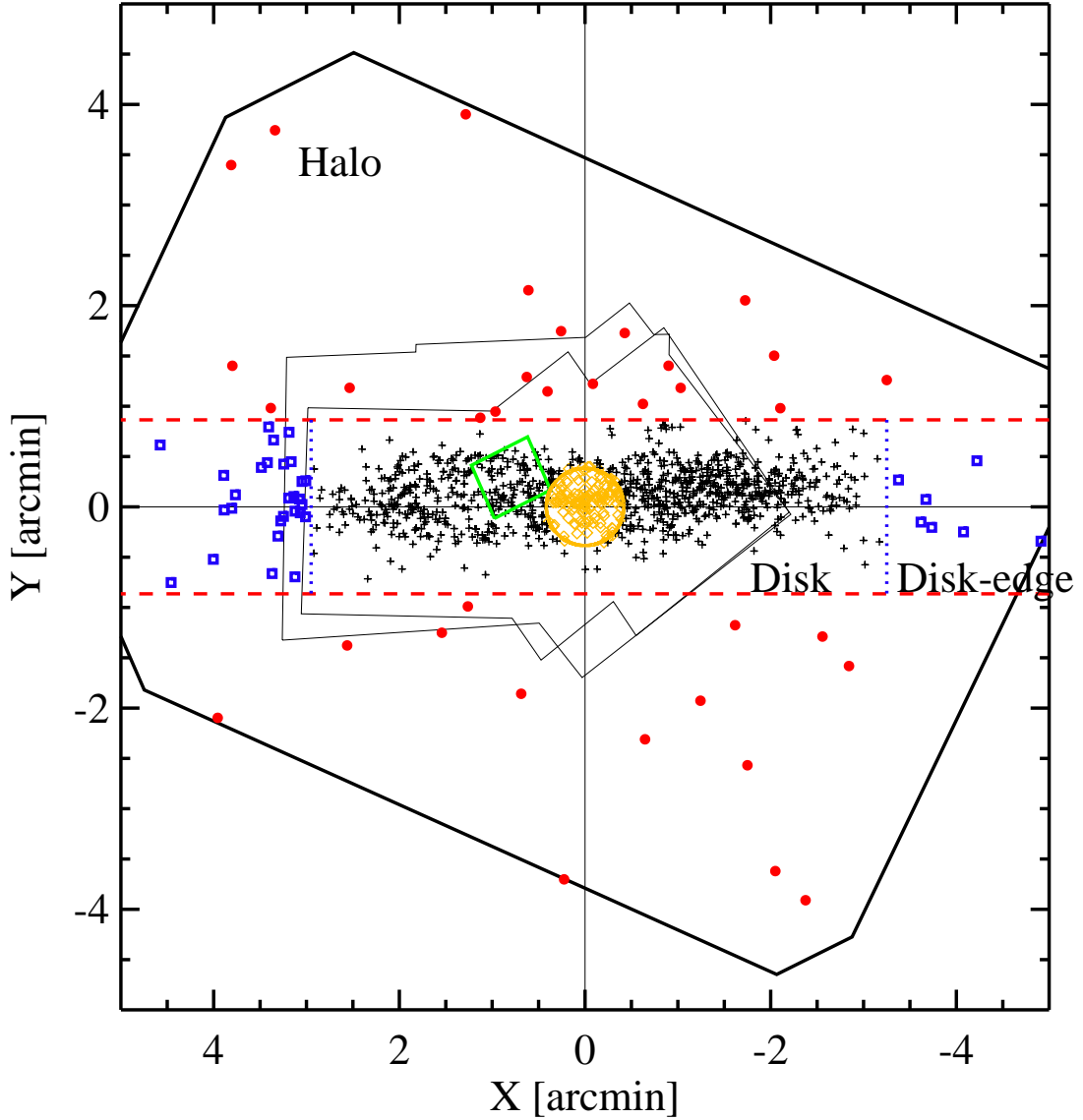} 
 \caption{Spatial distribution of the star clusters in M82. The large solid-line polygons 
 indicate the coverage of $HST$/ACS ($BVI$), $HST$/WFC3 UVIS ($U$), 
 and $HST$/WFC3 NIR ($YJ,H$) images. A large circle in the center represent the nuclear region with a radius of $25\arcsec$ (425 pc), 
 and a box left of the circle marks the region B ($35\arcsec \times 40\arcsec$, 595 pc $\times$ 680 pc). 
The horizontal dashed lines represent 
 the boundary between the disk region and the halo region. The dotted lines indicate the boundary between the disk region and the disk-edge region. 
 Crosses, diamonds, filled circles, and open squares represent the star clusters in the disk, nuclear region, halo, 
 and disk-edge region, respectively. One arcmin corresponds to one kpc.
\label{spa1}}
\end{figure}

Figure \,\ref{spa1} displays the spatial distribution of the star clusters in M82.
We defined the $X$ and $Y$ coordinates, which are projected distances along the major
and minor axes, respectively. $X$ and $Y$ increase towards the northeast and the northwest,
respectively, and are given by
$X=x'\sin($PA$)+y'\cos($PA$)$ and $Y=-x'\cos($PA$)+y'\sin($PA$)$
where $x'=(\alpha-\alpha_0) \cos(\delta)$, $y'=\delta-\delta_0$,  $\alpha_0$ and $\delta_0$ are the central coordinates of M82, and PA is the position angle for the major axis.

Several features are noted in this figure.
First, the most distinguishable feature is an elongated structure, showing that most star clusters are distributed on the disk:
they are mostly located within $ 1\arcmin$ from the major axis and $3\arcmin$ from the minor axis.
The distribution of the disk star clusters appears to show a discontinuity at $\sim 3\arcmin$ along the major axis, indicating the disk star cluster populations may include two kinds.
So we divided the disk area into two: the disk region ($-3 \arcmin \lesssim X \lesssim 3 \arcmin$, $-52 \arcsec \lesssim Y \lesssim 52 \arcsec$) and the disk-edge region ($\vert X \vert \gtrsim 3 \arcmin$, $-52 \arcsec \lesssim Y \lesssim 52 \arcsec$),
as marked in Figure \ref{spa1}.
Second, there are a small number of star clusters scattered far from the disk. These are probably old globular clusters belonging to the halo of M82.
Third, in the nuclear region that is known to have a recent starburst, we find 132 star clusters inside the circular area with a radius $25\arcsec$ (425 pc). \citet{May08} found 260 star clusters in the same region, including fainter clusters in $V$-band images as well as clusters in very complex regions.
Fourth, we find 80 star clusters in a  $35\arcsec \times 40\arcsec$  (595 pc $\times$ 680 pc) field centered on the region B, a famous post-starburst region.
We divide the entire field into four main regions for analysis:
the main disk region, the nuclear region, the disk edge region, and the halo region as marked in Figure \ref{spa1}.
Additionally, we select a sub region which is called region B for comparison with previous studies. 
This region is included in the disk region.

\begin{figure}
 \epsscale{1.0} \plotone{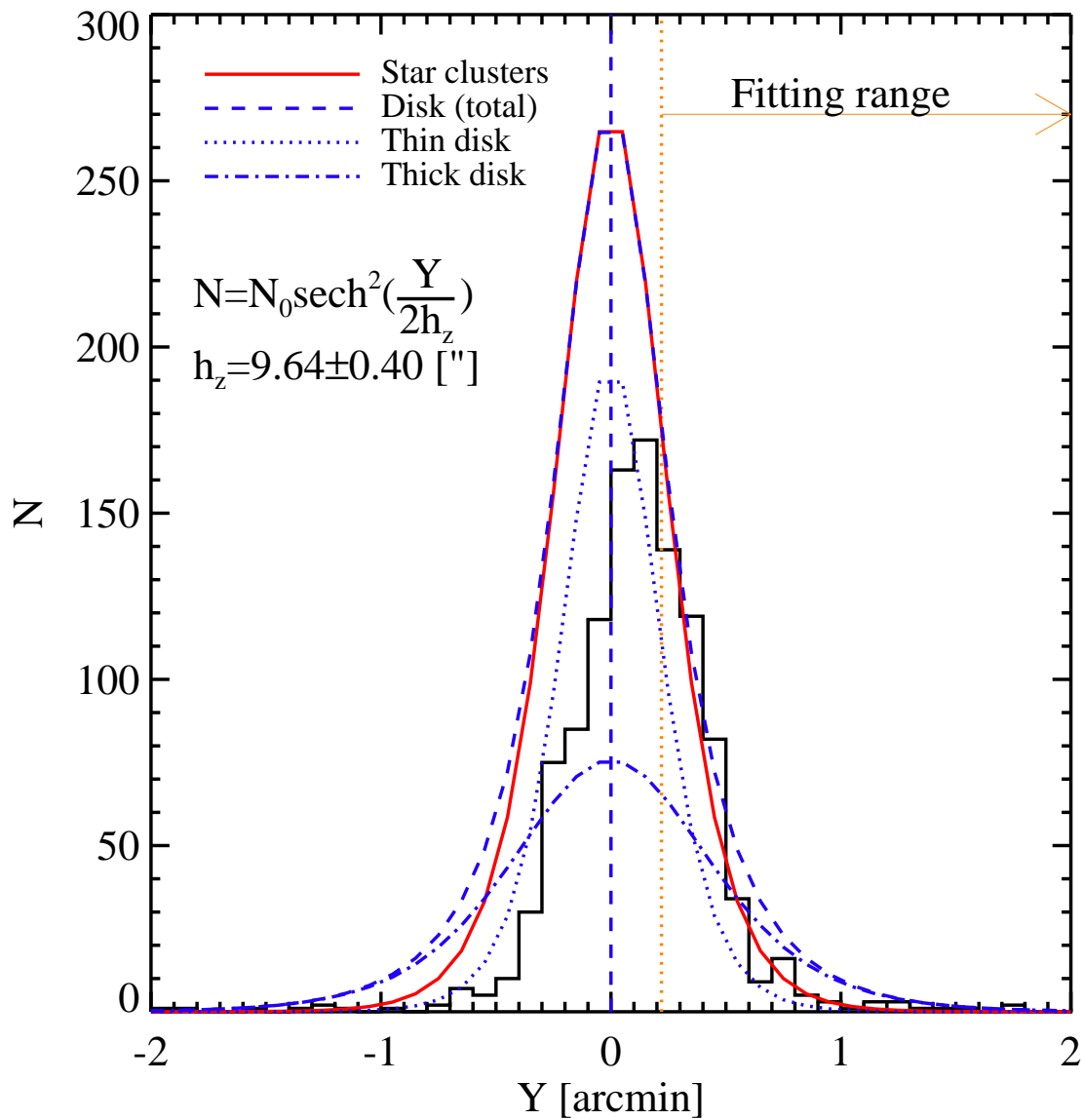} 
 \caption{The number density profile along the minor axis ($Y$) for the star clusters  in M82 (histogram), which is fit with a hyperbolic secant function (solid lines). 
 The vertical dotted line represent a lower limit for the fitting range. 
For comparison, fitting results for stellar light distribution are also shown for the thin disk (dotted line), thick disk (dot-dashed line), and sum of the two disks (dashed line). 
\label{verprof}}
\end{figure}

We derived a number density of star clusters as a function of distance from the major axis ($Y$) to investigate the vertical structure of the star cluster system,
as plotted in Figure \,\ref{verprof}.
The number density distribution appears to have  a Gaussian shape with asymmetry. 
The peak is shifted from the center to the northern-side ($Y>0$), and the number density of southern-side ($Y<0$) is smaller than that of northern-side.
This offset and asymmetry is considered to be due to the extinction and incompleteness of cluster detection. 
This shows that the star clusters in the southern-side suffer from more extinction than those in the northern-side, and that the southern-side ($Y<0$) is closer to us than the northern-side.

We fit the data with a hyperbolic secant function,  $N=N_0 $sech$^2 ( {{Y}\over{2h_z}} )$ where
$h_z$ is a scale height \citep{deG96}, plotting the result by the solid line in the figure.
The fitting is performed for the range of $Y>0.2\arcmin$ to avoid the incompleteness problem.
The vertical distribution of the star clusters is fit very well by the hyperbolic secant function
for $0.2\arcmin <Y<1\arcmin$. 
We also investigate the vertical structure of stellar populations using the $I$ band image. 
We 
sum counts with a range of $-50\arcsec < X < 50 \arcsec$ along the X-axis to derive a stellar light distribution along the Y-axis.
This distribution of stellar light shows asymmetry similar to that of star clusters.
We estimated the scale height of the vertical distribution of stellar light by fitting with a double hyperbolic secant function (for  thin and thick disks) for the range of $10\arcsec < Y < 100\arcsec$.
Derived scale heights of the thin and thick disks are $h_z({\rm thin, star})=8\arcsec .38\pm 0\arcsec .03$ ($\sim$143 $\pm$0.51 pc) and $h_z({\rm thick, star})=17\arcsec .29\pm 0\arcsec .05$ ($\sim$294 $\pm$0.85 pc), respectively. Thus the thick disk is twice thicker than the thin disk.

We derived a value for $h_z$ of the star cluster distribution, 
$9\arcsec.64\pm0\arcsec.40$ ($164\pm7$ pc).
This value is a little larger than that of the thin disk derived from stellar light, but this distribution includes the intermediate-age and old star clusters that are regarded as thick disk or halo populations.
The genuine $h_z$ of the star clusters in the disk could be similar to that of stellar thin disk.
It suggests that most star clusters are distributed similarly to stellar populations in the thin disk.

It is also noted that there are several star clusters at $Y>1\arcmin$ above the fit function.
 These clusters may be a separate population from the disk population.
They are considered to  belong to the halo of M82.
We defined the halo region by  $|Y| > 3 h_z ({\rm thick})$  (= $51 \arcsec.87$ = 882 pc) derived from the  $I$-band
stellar luminosity profile for analysis, as marked in Figure \ref{spa1}.

\subsection{Photometric Properties of Star Clusters}

\subsubsection{Color-Magnitude Diagrams and Color Distribution of the Star Clusters}

\begin{figure}
 \epsscale{1.0} \plotone{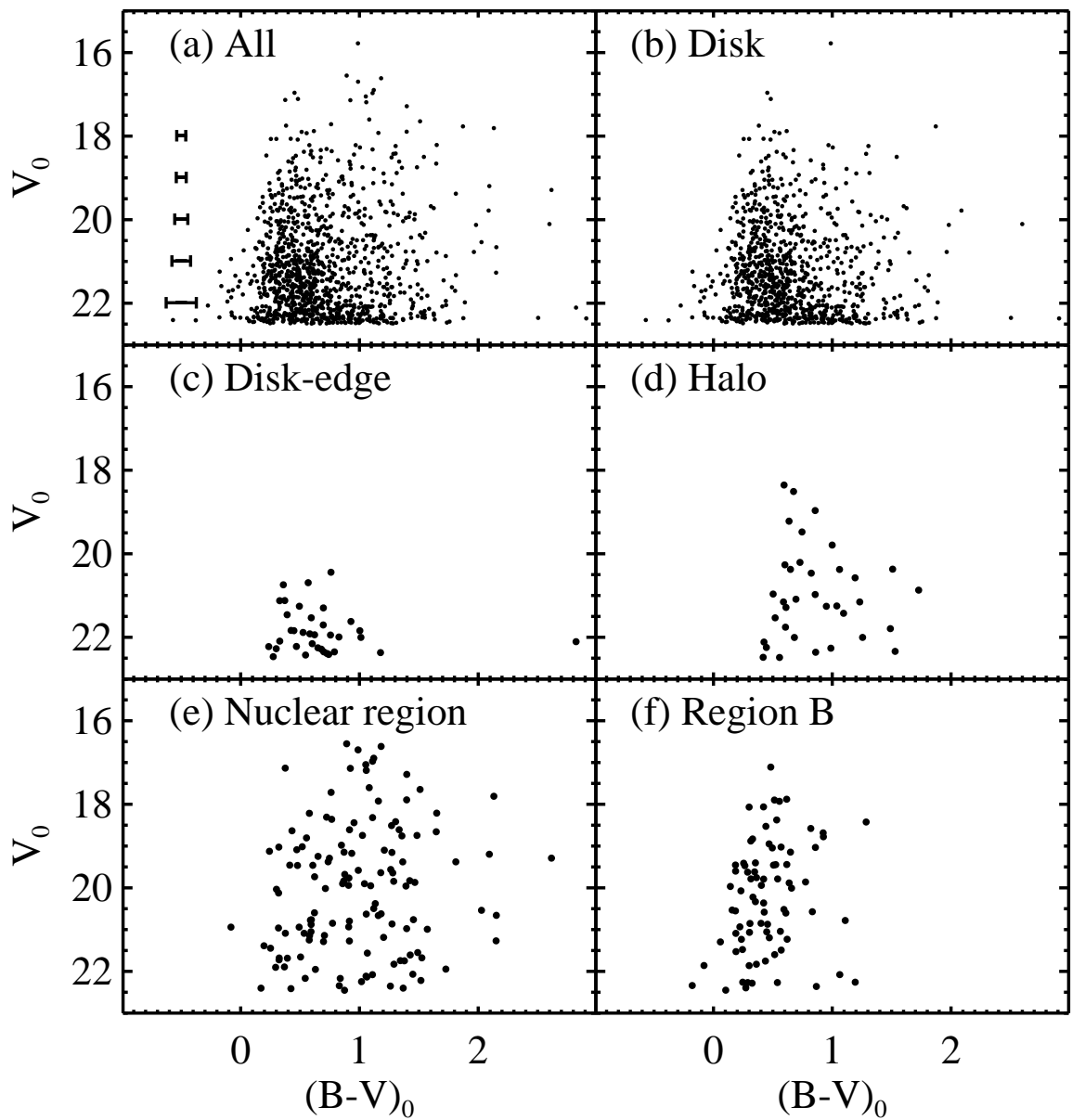} 
 \caption{Color-magnitude diagrams of the star clusters in M82:
 (a) the entire region, (b) the disk region, (c) the disk-edge region, (d) the halo region,
 (e) the nuclear region, and (f) region B.
 The bars at the left side in (a) show mean $(B-V)$ errors for magnitude bins. 
\label{vbvcmd}}
\end{figure}

\begin{figure}
 \epsscale{1.0} \plotone{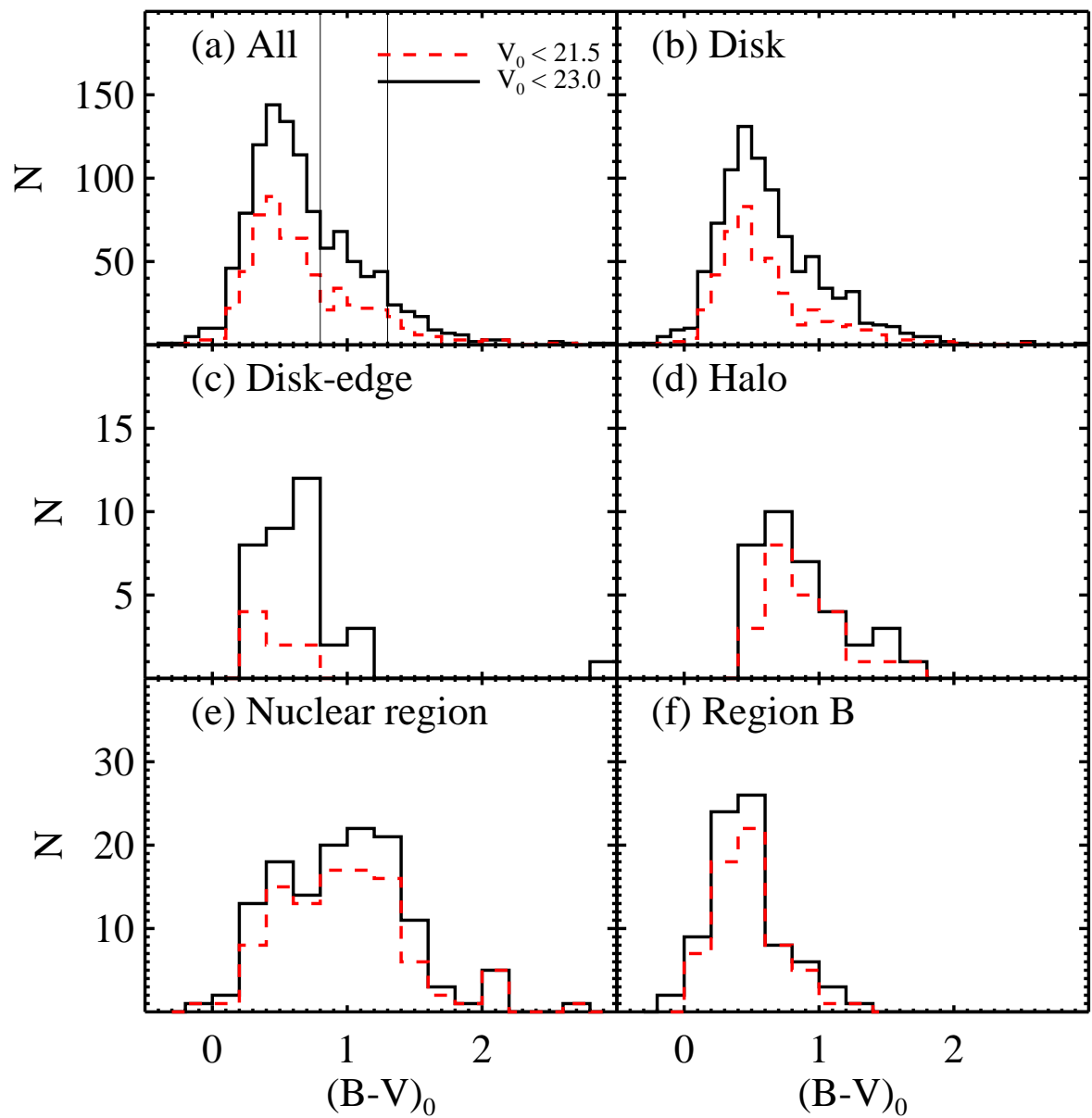} 
 \caption{$(B-V)$ color histograms of all star clusters (solid line histogram) and bright star clusters with
 $V_0<21.5$ mag (dashed line histograms) in M82: (a) the entire region, (b) the disk region, (c) the disk-edge region, (d) the halo region,
 (e) the nuclear region, and (f) region B.
Vertical lines in the panel (a) represent the boundary of three groups: blue, intermediate-color, and red clusters.
\label{colhist}}
\end{figure}

Figure \,\ref{vbvcmd} displays $V_0 - (B-V)_0$ color-magnitude diagrams (CMDs) of the star clusters in M82, 
 and Figure \,\ref{colhist} shows corresponding $(B-V)_0$ color histograms (for all star clusters and
 only bright star clusters with $V_0<21.5$ mag):
 (a) the entire area, (b) the disk region, (c) the disk-edge region, (d) the halo region,
 (e) the nuclear region, and (f) the region B. 
 Here magnitudes and colors represent the
 values dereddened according to the foreground reddening.
Distinguishable features are as follows.
First, the star clusters in M82 are dominated by the star clusters in the disk.
They are composed of a dominant blue population
with a color peak at $(B-V)_0 \approx 0.45$, and a weaker redder population
that are mostly redder than $(B-V)_0 \approx 1.0$.
We divide the cluster sample according to their color range:
$(B-V)_0 \leqq 0.8$ (blue clusters),  $0.8<(B-V)_0 \leqq 1.3$ (intermediate color clusters), and $(B-V)_0  >1.3$ (red clusters) for further analyses.
Second, the CMDs and color distributions vary depending on the region.
Third, the star clusters in the disk-edge region and the region B are mostly blue.
Fourth, the brightest star clusters in the disk-edge region, $V_0 \approx 20.5$ mag, are 
much fainter than those in other regions.
Fifth, the star clusters in the nuclear region include
a large number of brighter star clusters compared with other regions,
and they show a large scatter in color, indicating a presence of high differential reddening.
Finally, the star clusters in the halo region are mostly red with $0.4<(B-V)_0 \lesssim 1.2$.
These may be old globular clusters, showing the existence of an old halo in M82.
A few of them have very red colors  $(B-V)_0 > 1.4$, indicating that some of them may be
reddened star clusters or background galaxies.

\subsubsection{Spatial Distribution of the Star Clusters with Different Colors}

\begin{figure}
 \epsscale{1.0} \plotone{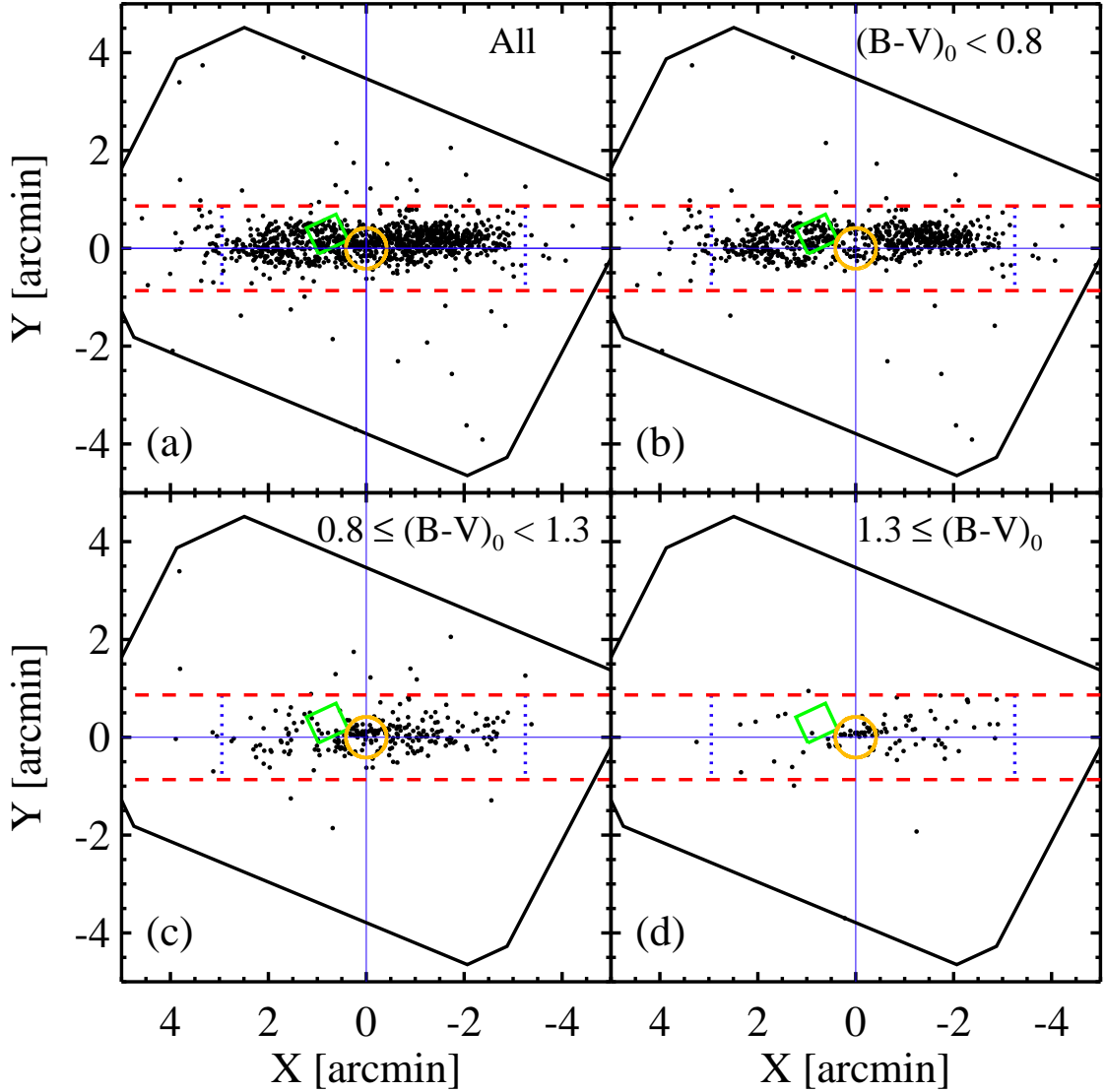} 
 \caption{Spatial distributions of the star clusters in M82 with different color ranges:
   (a) all, (b) blue clusters ($(B-V)_0 < 0.8$), (c)  intermediate color clusters ($0.8\leqq (B-V)_0 <1.3$), and (d) red clusters ($1.3 \leqq (B-V)_0$).
 The large solid-line box indicates the coverage of 
 $HST$/ACS ($BVI$).
The dashed lines represent the boundary between the disk  and halo regions,
and the dotted lines represent the boundary between the disk  and disk-edge regions. 
\label{spa2}}
\end{figure}

The spatial distributions of the star clusters with different color ranges are shown in Figure \,\ref{spa2}.
The blue star clusters ($(B-V)_0 < 0.8$) are mostly distributed in the disk, and
several of them are found in the halo region. 
There are 69 and 48 blue star clusters in the region B and nuclear region, respectively.
The intermediate-color clusters ($0.8 \leqq (B-V)_0 < 1.3$) are also found mostly in the disk region and a few are in the halo region.
Interestingly the distribution in the left is much wider than that in the right. 
It could be a result due to inhomogeneous extinction. 
A more detailed description is given in Section 3.6.
Ten and 52 intermediate-color clusters are found in the region B and
nuclear region, respectively. The red star clusters ($(B-V)_0 \geqq 1.3$) are
mostly located in the disk, especially in the right  side of the disk.
Only one red cluster is found in the region B, while 32 red clusters are detected in the nuclear region.
Interestingly the red clusters are located along a narrow linear structure.
We checked the $YJ$ and $H$ band images, and found the existence of dust lanes in the nuclear region. 
These dust lanes obscure the star clusters in the nuclear region, and the star clusters in the edge of the dust lanes are only shown as a narrow linear structure.
The star clusters in the region B are usually blue, while more than half of star clusters in the nuclear region are red.
It could be caused by the different age of star clusters or differential reddening.
The region B is regarded as a window with less extinction and the nuclear region is usually considered as a high extinction region, so that the red colors of most clusters in the nuclear region 
may be a result of internal reddening.
The linear distribution of the red star clusters in the nuclear region
implies that there is a narrow window in the nuclear region
or the clusters are formed in such a location.

\subsubsection{Color-Color Diagrams of the Star Clusters}

\begin{figure}
 \epsscale{0.8} \plotone{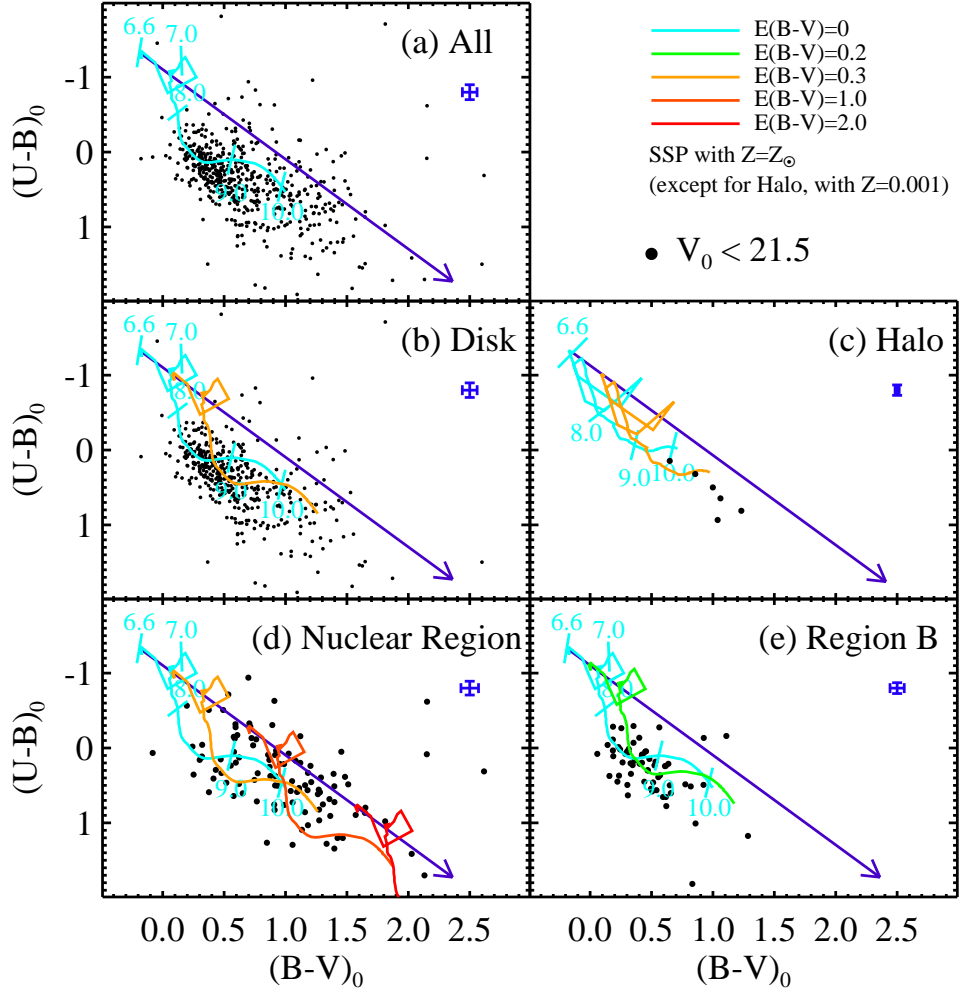} 
 \caption{($U-B$)$_0$ -- ($B-V$)$_0$ color-color diagrams of the bright star clusters with $V_0<21.5$ mag in  M82: (a) the entire region, (b) the disk region, (c) the halo region, (d) the nuclear region, and (e) 
 region B. The solid curved lines show the 
 simple stellar population model for $Z=Z_{\odot}$ (except for the halo, $Z=0.001$)
 given  by \citet{Mar08}, shifted according to $E(B-V)=0.0$, 0.2, 0.3, 1.0, and 2.0. 
 The arrows represent reddening vectors   with $A_V =9$. Error bars at the right-upper corner of all panels represent median values of photometric errors.
\label{ubvccd}}
\end{figure}

Figure \,\ref{ubvccd} displays the $(U-B)_0 -(B-V)_0$ color-color diagrams (CCDs) of the bright star clusters with $V_0 <21.5$ mag in M82:
(a) the entire area covered by $U$ images, (b) the disk region,  (c) the halo region,
 (d) the nuclear region, and (e) the region B.
 Note that the halo region is partially covered by $U$ images so that only six objects are plotted.
We also plotted  
the SSP models with solar metallicity from the Padova group \citep{Mar08}.
Several features are noted in Figure \,\ref{ubvccd}.
First, the disk star clusters are mostly in the region for $8< {\rm log (age)} <10$ and $E(B-V)\approx 0.3$.
Second, the halo star clusters are old (log(age)$>9$).
Third, the star clusters in the nuclear region are mostly $8< {\rm log (age)} <9$ and $0.1<E(B-V)\lesssim 1.0$. However, there are some clusters that are severely reddened ($0.2<E(B-V)\lesssim 2.0 $ or more).
Fourth, the region B star clusters are mostly located in the region for $8< {\rm log (age)} <9$ and $E(B-V)\approx 0.2$.

\subsection{Luminosity Functions of the Star Clusters}

\begin{figure}
 \epsscale{1.0} \plotone{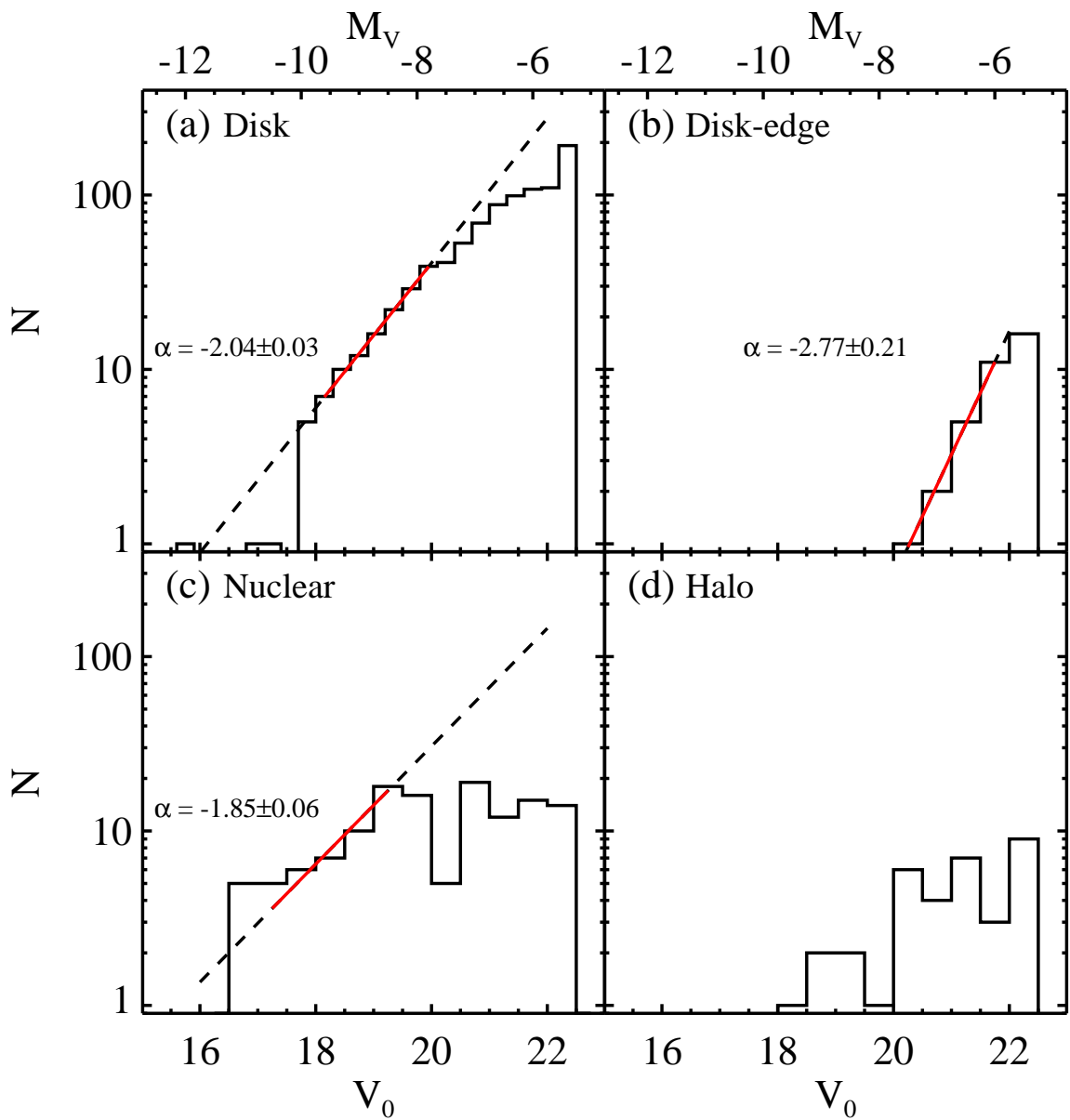} 
 \caption{Luminosity functions of the star clusters in the  
 disk (a), disk-edge (b), nuclear (c), and halo (d) regions of M82.
 The thick solid lines in (a), (b), and (c) show a power-law fit to the luminosity function 
 of the star clusters over the $18.0 \geq V_0 \geq 20$, $20.0 \geq V_0 \geq 22$, and $17.5 \geq V_0 \geq 19.5$, respectively (the thick dashed lines
 represent an extrapolation). 
\label{lf}}
\end{figure}

The luminosity function  of the star clusters in M82 is shown in Figure \,\ref{lf}.
The luminosity function of the star clusters with $-10 < M_V < -8$ in the disk is fit well by a power-law function, $dN/dL \propto L^{\alpha}$ with $\alpha=-2.04\pm0.03$.
This value is similar to that of young star clusters in other late-type galaxies (e.g. \citealp{Whi99,Hwa08,San10}).
\citet{Ann11} suggested that the slope of the luminosity function of the star clusters in the spiral galaxies becomes steeper when the luminosity of the fitting range is increased. The slope of the luminosity function for the star clusters in the disk region in M82 also follows this trend. The faint end in the luminosity function for the star clusters in the disk region becomes flatter at $V_0 \sim 21$ mag, which is a result due to incompleteness. This flattening is also seen in the luminosity function of M82 star clusters derived from the same data by \citet{May08}.

The luminosity function of the star clusters with $-8<M_V <-6$ in the disk-edge region is also fit by a power-law function with $\alpha=-2.77\pm0.21$, which is a little steeper than that of the star clusters in the disk region.
The brightest star cluster in the disk-edge region is much fainter than those in other regions, as mentioned at Section 3.3.1. 
This result suggests two conflicting interpretations that the bright clusters could not be formed in this region or bright clusters were disrupted in this region. The former is more probable, considering that more massive clusters form in denser environments.
The luminosity function of the star clusters in the nuclear region is fit by a power-law function with a shallower slope of $\alpha=-1.85\pm0.06$. However, it is noted that there is an excess in the bright end of the luminosity function for the nuclear region.
It suggests that the starburst phenomenon in the nuclear region could form very luminous star clusters, but we should deal carefully with this result because of extinction and incompleteness.
The luminosity function of the star clusters in the halo region cannot be fit by a power-law function. 
It is noted that there are bright clusters with $M_V < -8$ in the halo region that are not found in the disk-edge region, while the numbers of the star clusters in the halo and disk-edge regions are similar.

\subsection{Size Distribution of the Star Clusters}

\begin{figure}
 \epsscale{1.0} \plotone{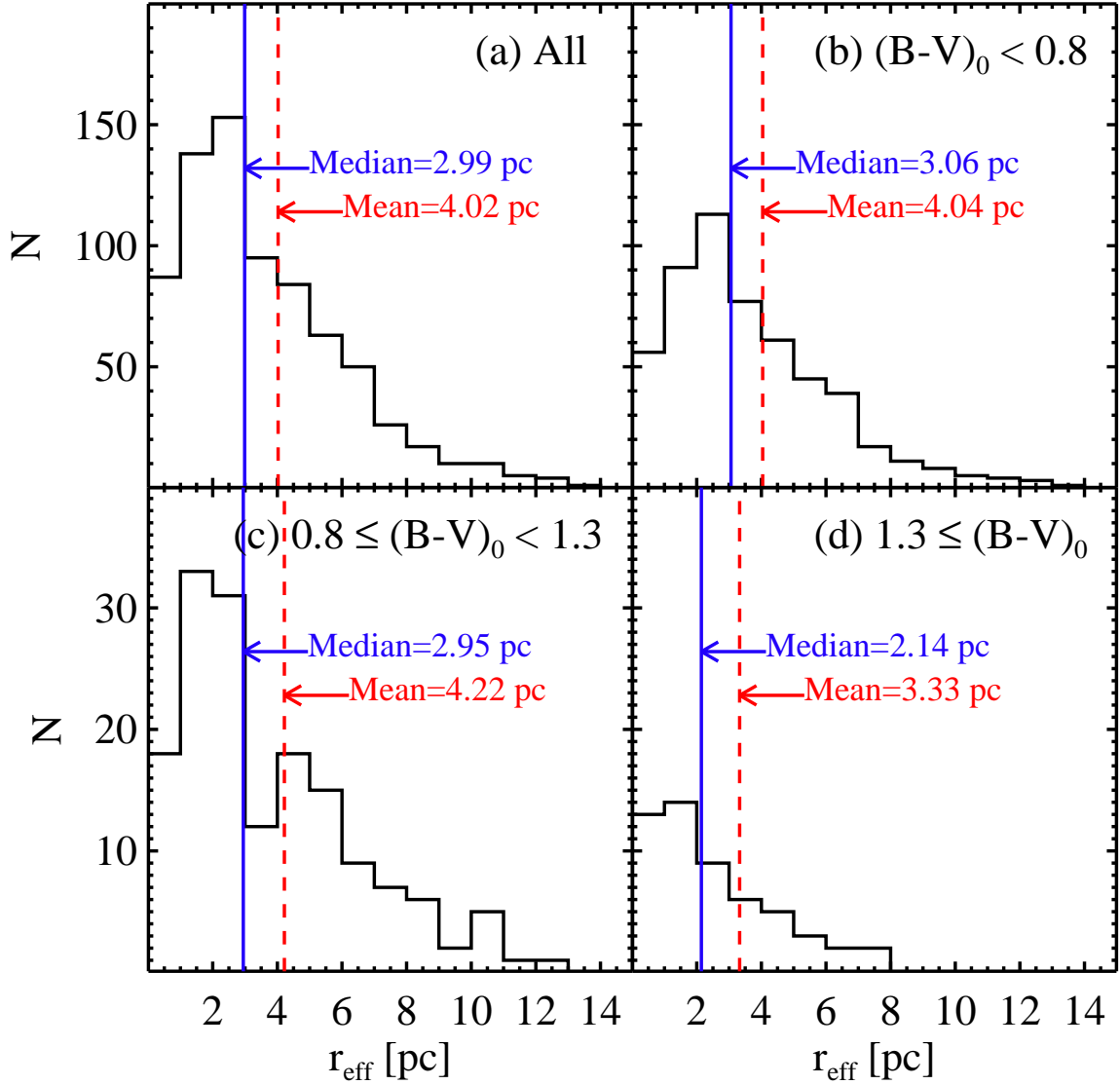} 
 \caption{Size distributions of the star clusters in M82:
  (a) all, (b) blue clusters ($(B-V)_0 < 0.8$), (c)  intermediate color clusters ($0.8\leqq (B-V)_0 <1.3$), and (d) red clusters ($1.3 \leqq (B-V)_0$).
 The vertical solid and dashed lines 
 indicate the median and mean values of effective radii, respectively.
\label{size1}}
\end{figure}

Figure \,\ref{size1} displays the distributions of effective radii of the star clusters in M82:
(a) all, (b) the blue star clusters, (c) the intermediate-color star clusters, and (d) the red star clusters.
The star clusters have sizes of $0 < r_{\rm eff} \lesssim 14$ pc, but mostly smaller than 8 pc.
The size distribution shows a peak at  $r_{\rm eff} \sim 2$ pc, and it smoothly decreases as the star cluster size increases.
The median and mean values of the star cluster sizes are
2.99 pc and 4.02 pc for all star clusters,
3.06 pc and 4.04 pc for the blue star clusters,
2.95 pc and 4.22 pc for the intermediate-color star clusters, and
2.14 pc and 3.33 pc for the red star clusters.
The mean value of M82 star cluster sizes derived in this study is consistent with the peak value 
$r_{\rm eff}=$3.8 pc given by \citet{May08}.
There are no significant differences of sizes between the blue and intermediate color star clusters (mean values are 4.04 pc and 4.22 pc, respectively). However, the sizes of the red clusters (a mean value of 3.33 pc) are a little smaller than the blue clusters (4.04 pc), but their standard deviations are too large ($\sim5$) to insist the difference.

\begin{figure}
 \epsscale{1.0} \plotone{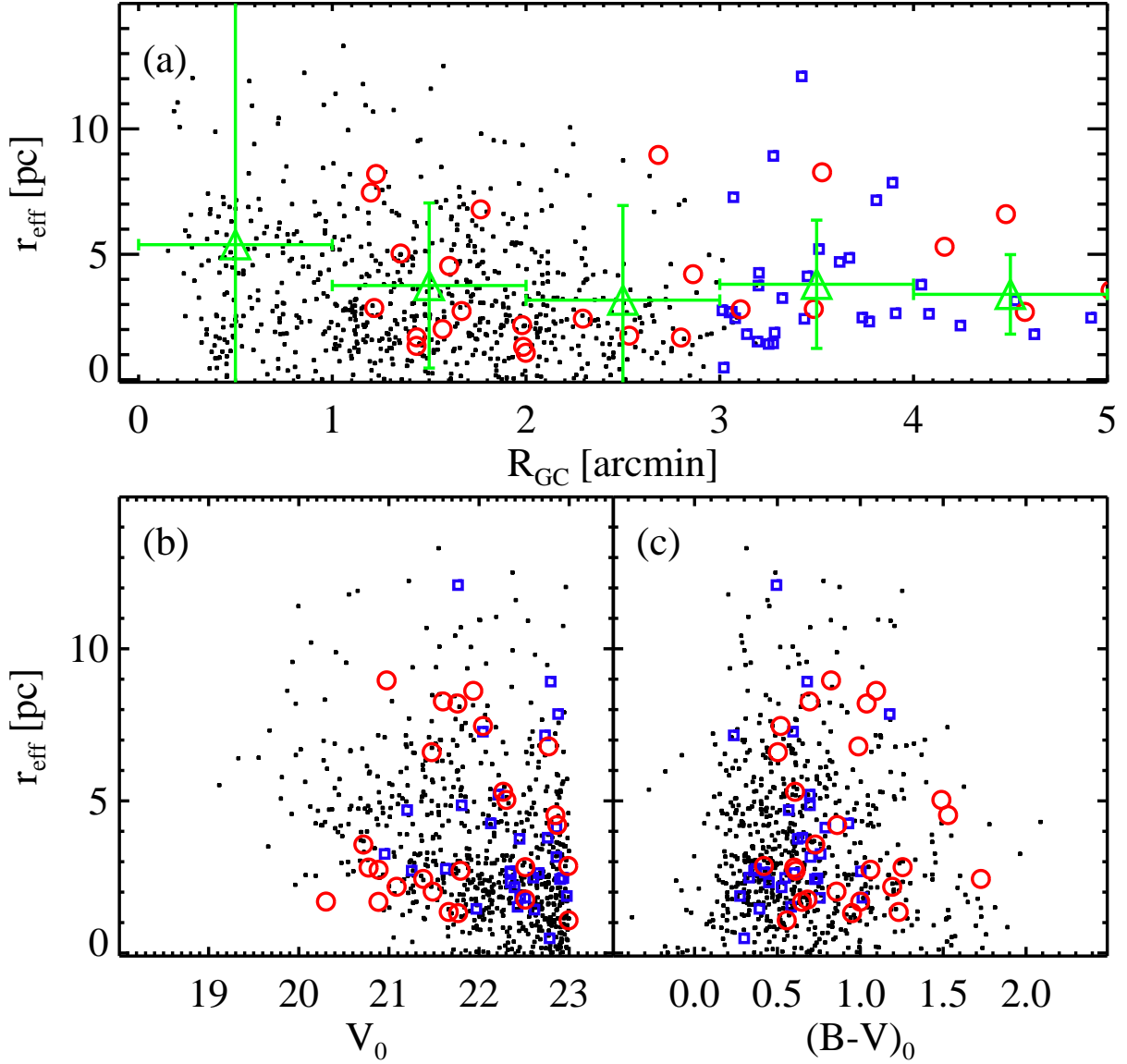} 
 \caption{Size versus galactocentric radii, $R_{GC}$ (a),
 $V_0$ (b), and $(B-V)$ (c) for the star clusters in M82.
 Small filled circles, large open circles, and open squares represent the star clusters with good-fit sizes ($\chi^2<10$) in the disk, halo, and disk-edge region, respectively. 
Triangles with error bars in (a) represent mean values of the star cluster sizes in each bin. The error bars of $X$ axis direction and $Y$ axis direction show radial bin sizes and standard deviations of mean sizes, respectively.
\label{size2}}
\end{figure}

Figure \,\ref{size2}(a) displays the sizes of the star clusters versus
the galactocentric distance ($R_{GC}$).
There is seen no significant trend of the sizes depending on $R_{GC}$.
However it is noted that large star clusters with $r_{\rm eff} >10$ pc are found mostly in the inner region of the disk at $R_{GC}<1\arcmin.5$.
Figure \,\ref{size2}(b) and (c) displays the sizes of the star clusters versus $V$-band magnitudes and  $(B-V)$ colors, respectively.
Large star clusters with $r_{\rm eff} > 10$ pc are mostly fainter than $V \sim 20$ mag, while small star clusters with  $r_{\rm eff} < 10$ pc have a wide range of magnitudes.
The $(B-V)$ color of the star clusters does not show any significant
dependence on the size of the star cluster.

The peak value for M82 star clusters, 2 pc, is similar to that of the star
clusters in other galaxies: e.g., 2.27 pc for M51 \citep{Hwa08}, $\sim2.3$ pc for M83 \citep{Bas12}, and $\sim1.5$ pc for M31 \citep{Van09}. A relation between cluster radii and galactocentric distance is found for the star clusters of M51 and M83 \citep{Sch07,Hwa08,Bas12}.
The star clusters in the outer region of those galaxies have larger sizes than those in the inner region of those galaxies.
However, the star clusters in M82 do not follow this trend. 
We also checked the relation between the sizes and distances from the major axis, finding no significant trend.

\subsection{Age, Extinction, and Mass of the Star Clusters}

\begin{figure}
 \epsscale{1.0} \plotone{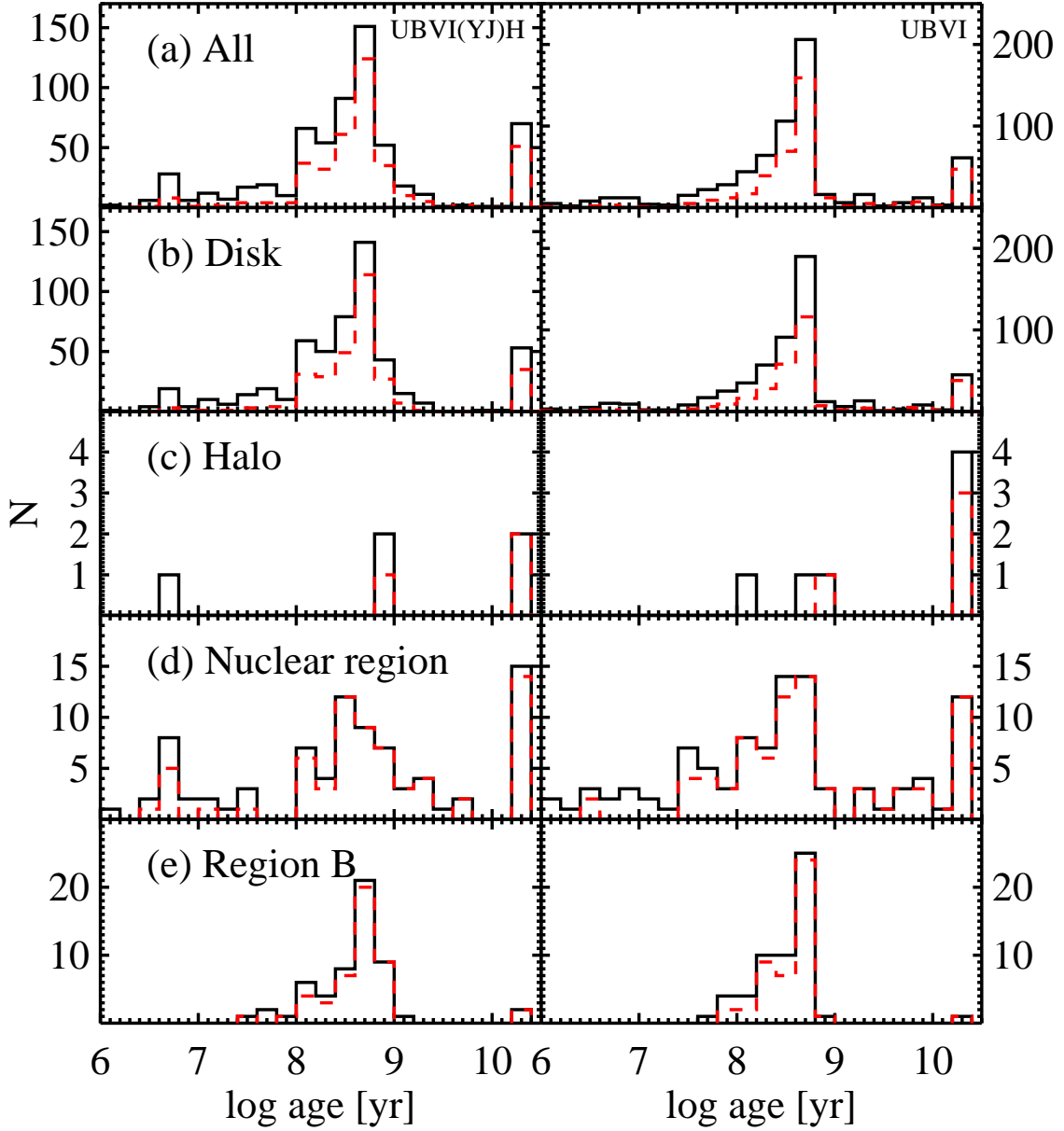} 
 \caption{Age distributions of the star clusters in M82: 
(a) the entire region, (b) the disk region,  (c) the halo region,
 (d) the nuclear region, and (e) region B.
 The solid and dashed lines indicate all star clusters and bright star clusters with 
 $V_0<21.5$ mag, respectively. 
 The left  and right panels show the ages of the  star cluster derived from $UBVI(YJ)H$ 
 and $UBVI$ , respectively.
\label{age1}}
\end{figure}

Figure \,\ref{age1} displays the age distribution of $\sim$630 star clusters in M82: (a) the entire region covered by the $U$-band images, (b) the disk region, (c) the halo region,
 (d) the nuclear region, and (e) region B.
The ages of the star clusters are derived from SED fitting with two filter combinations, $UBVI(YJ)H$ (left panels) and  $UBVI$ (right panels).
Several distinguishable features are noted in Figure \,\ref{age1}(a).
First, there is a dominant population of
intermediate-age star clusters having ages from 100 Myr to 1 Gyr
with a  peak at $\sim500$ Myr (${\rm log(age)}=8.7$).  
This is seen in both  $UBVI(YJ)H$ fit results and  $UBVI$ fit results.
There appears to be a separate weaker peak at $\sim125$ Myr (${\rm log(age)}=8.1$).
This is better seen in the left panel than in the right panel.
Second, there is a small number of young population with $<10$ Myr.
Third, there is a sizable population of old star clusters with $>10$ Gyr.

Figure \,\ref{age1}(b) to (e) show the variation of the age distribution of the star clusters
depending on their positions.
First, the age distribution of the disk star clusters is similar to that for all star clusters because
the entire cluster system in M82 is  dominated by the disk clusters.
Second, the number of the halo star clusters with age estimates is small, but they are mostly
older than 1 Gyr.
Third, the star clusters in the nuclear region show three populations: a major intermediate-age population
with a peak value at  $\sim400$ Myr (${\rm log(age)}=8.6$), a relatively small number of very young population with $<10$ Myr as well as old population with $>10$ Gyr.
Fourth, the star clusters in the region B are mostly intermediate-age clusters.

\begin{figure}
 \epsscale{1.0} \plotone{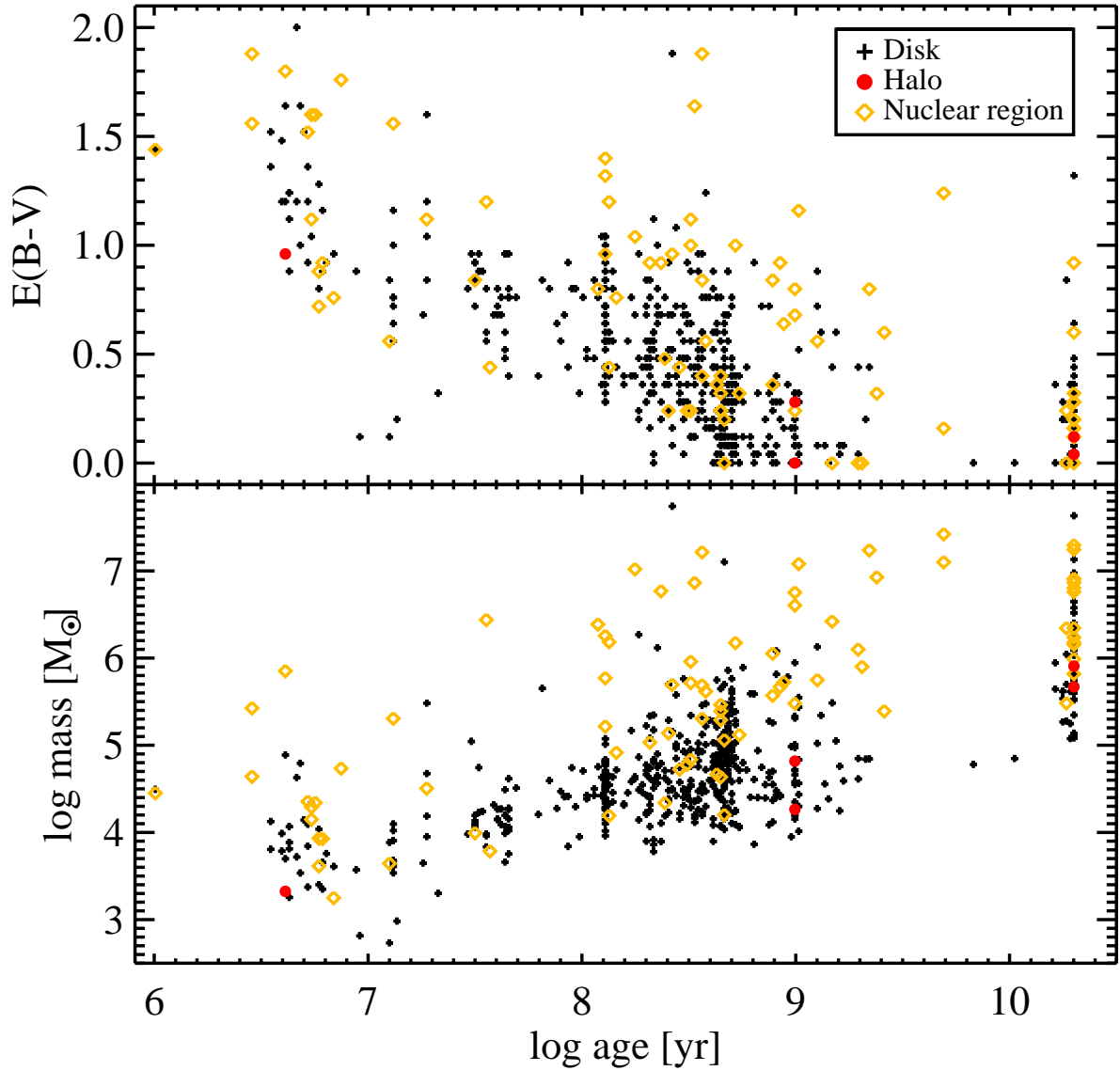} 
 \caption{$E(B-V)$ versus log(age) (a) and log $M/M_{\odot}$ versus log(age) (b).
 Crosses, filled circles, and diamonds represent the star clusters in the disk, halo, and nuclear regions, respectively.
\label{mass1}}
\end{figure}

Figure \,\ref{mass1} displays $E(B-V)$ and masses versus ages of the star clusters in the halo, disk, and nuclear regions.
$E(B-V)$ values of the star clusters in the halo region are mostly small with $E(B-V)<0.3$.
$E(B-V)$ values of the star cluster in the disk and nuclear regions are mostly distributed in the range of $0 \lesssim E(B-V) \lesssim 1.0$.
The reddening value appears to increase as cluster ages decrease.
In particular, for the young star clusters in the disk and nuclear regions with age $<10$ Myr the reddening values are $0.8 \lesssim E(B-V) \lesssim 2.0$.
This indicates that young star clusters are usually located in dusty area suffering from high extinction, or that ages of these clusters might be under-estimated due to overestimation of extinction.

The masses of the star clusters in M82 have a range with
$10^3 M_{\odot} \lesssim M \lesssim 10^7 M_{\odot}$, but mostly
$10^4 M_{\odot} \lesssim M \lesssim 10^6 M_{\odot}$.
The lower boundary of the cluster masses is defined by the survey limit of this study.
Most massive star clusters with $M\gtrsim 10^6 M_{\odot}$ are located in the nuclear region.
Several clusters with ${\rm log(age)} \gtrsim 8$ have very high masses with
$M \sim 10^7 M_{\odot}$, but it should be treated with caution because over estimation of the extinctions could lead to high mass.
Young star clusters with ages  $<10$ Myr have mostly low mass
$M < \sim 10^{4.5} M_{\odot}$.
Old star clusters are mostly
massive with $10^{5.2} M_{\odot} \lesssim M\lesssim 10^{6.5} M_{\odot}$.

\begin{figure}
 \epsscale{1.0} \plotone{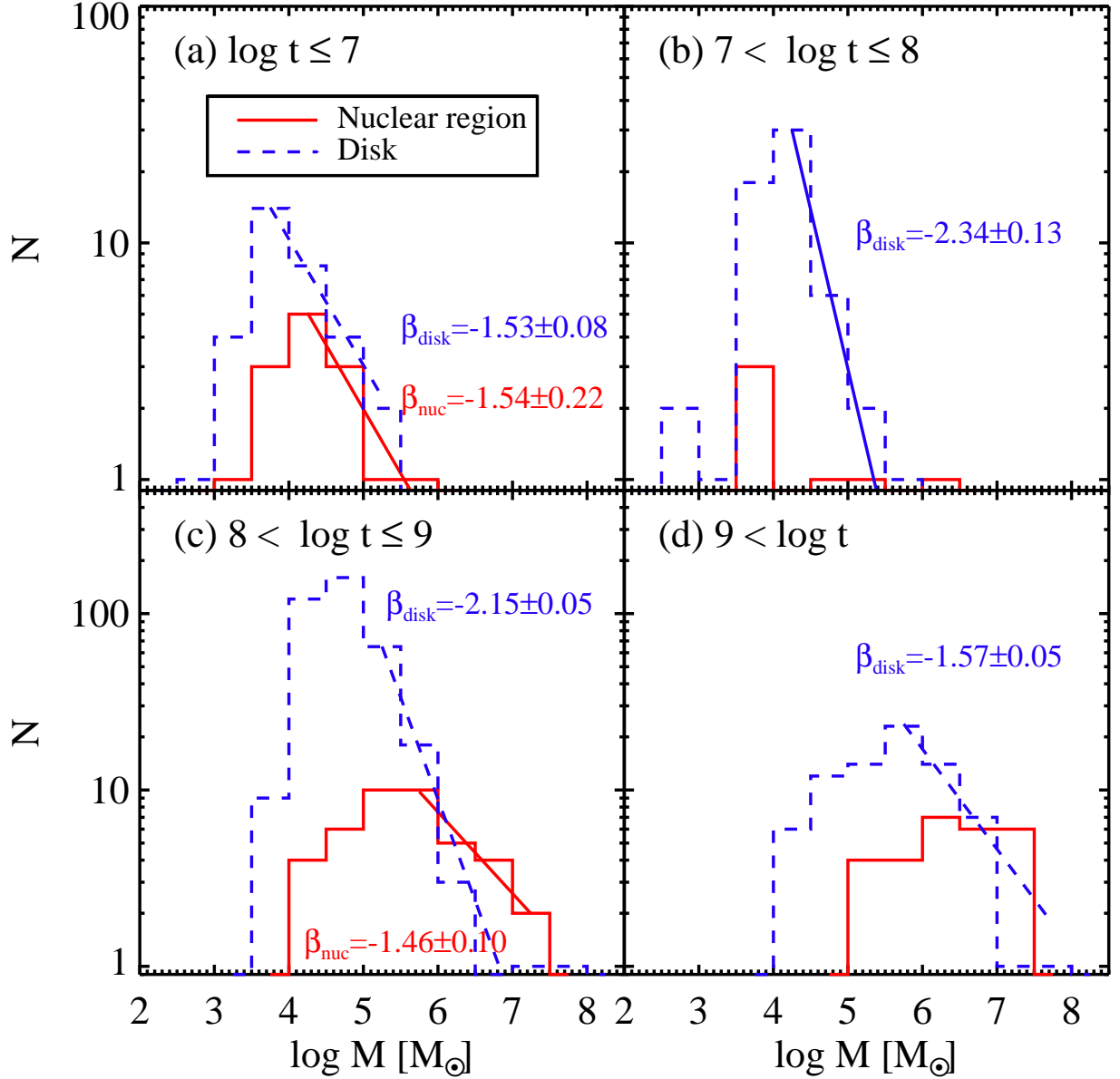} 
 \caption{Mass functions of the star clusters in the nuclear region (solid lines) and disk (dashed lines) with different ages:
 (a) log(age)$<7$,  (b) $7 <$ log(age) $\leqq 8$, (c) $8 <$ log(age) $\leqq 9$, and (d) $9 <$ log(age).
 Linear lines represent power law fits for the upper mass range.
\label{mass2}}
\end{figure}

Figure \ref{mass2} displays the mass functions 
of the star clusters in the disk and nuclear region with different ages. 
There are several noticeable features as follows.
(1) Upper parts of the mass functions of the young star clusters with ${\rm log (age)} \le 9$ in the disk and nuclear region mostly follow power-law functions ($N(m)dm \propto m^{\beta}$). 
(2) The mass of the youngest star clusters with ${\rm log (age)} \le 7$ in both the disk and nuclear regions have similar slopes of power-law functions with $\beta_{disk}=-1.53\pm0.08$ and $\beta_{nuc}=-1.54\pm0.22$  (fitting ranges of $10^{3.5} M_{\odot} < M < 10^{5.5} M_{\odot}$ for the star clusters in the disk region and $10^{4.0} M_{\odot} < M < 10^{6.0} M_{\odot}$ for the star clusters in the nuclear region). 
(3) The mass functions of the intermediate-age star clusters with $8< {\rm log (age)} \le 9$ in both regions have significantly different slopes of power-law functions.
(4) The mass functions of the old star clusters in the disk do not follow a power-law function.

It is known that the mass function of the young star clusters in most disk galaxies follows a power-law functions with
$\beta\sim-2$ (e.g. for M51 young star clusters (${\rm log(age)}<7.0$), $\beta=-2.23\pm0.34$ with $10^{3.6} M_{\odot}\lesssim  M \lesssim 10^{4.1} M_{\odot}$ \citep{Hwa10}; and for the young star clusters in Antennae galaxies (${\rm log(age)}<7.0$), $\beta=-2.14\pm0.04$ with $10^4M_{\odot}\lesssim  M \lesssim 10^6M_{\odot}$ \citep{Whi10}, but the mass functions of the  youngest star clusters in M82 have a shallower slope with $\beta\sim-1.5$.
It may be understood as a shallower initial cluster mass function (ICMF) of starburst galaxies, but the number of the young star clusters is too small to interpret this result as a top heavy ICMF.
It is also noted that the mass function of the intermediate-age star clusters in the nuclear region is shallower with $\beta =-1.46\pm0.10$ than that of typical young clusters, as shown in Figure \ref{mass2}(c).
However, this mass function can be interpreted as a result of star cluster disruption or incomplete survey.

There are several extremely massive star clusters ($M\gtrsim10^{6.5}M_{\odot}$) in the nuclear region with $8<{\rm log(age)}\leq9$.
They are more massive than typical globular clusters.
The existence of the high mass star clusters could be interpreted as a result of a starburst phenomena of M82 at the epoch of the galaxy interaction.
According to the spectroscopic study by \citet{Lan08}, two luminous star clusters have dynamical masses of  $10^6-10^{6.6}M_{\odot}$, which is slightly less massive than our results suggest ($\sim 10^7 M_{\odot}$), but they are still super massive.

\begin{figure}
 \epsscale{1.0} \plotone{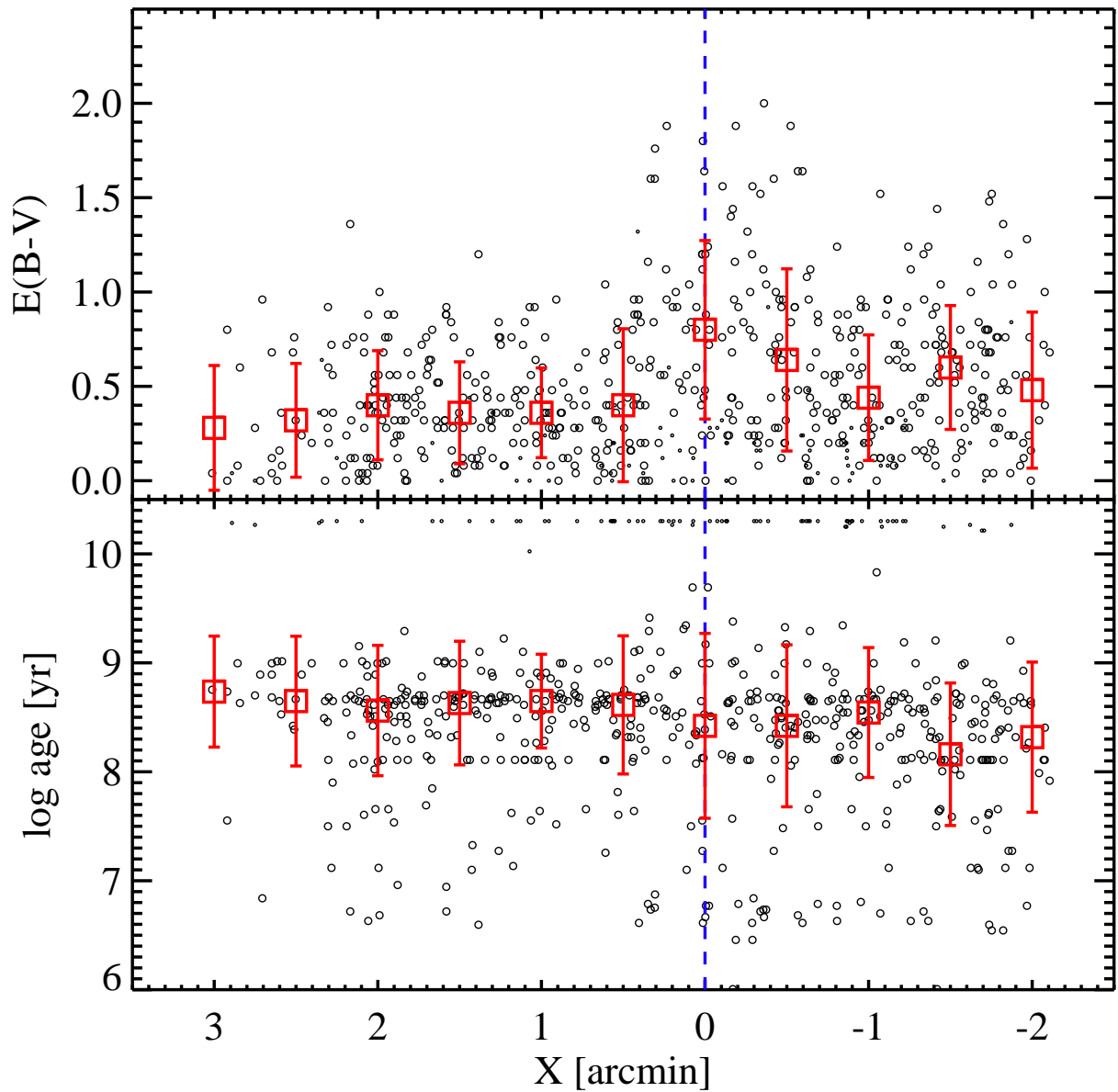} 
 \caption{ $E(B-V)$  (upper panel) and ages (lower panel) versus $X$ for the M82 star clusters. The squares with error bar represent the mean values of ages and $E(B-V)$ for the star clusters with log(age)$<10$. 
Data with log(age)=10.3 represent the star clusters for which fitting failed and that were assigned very old ages.
\label{ebvageX}}
\end{figure}

\begin{figure}
 \epsscale{1.0} \plotone{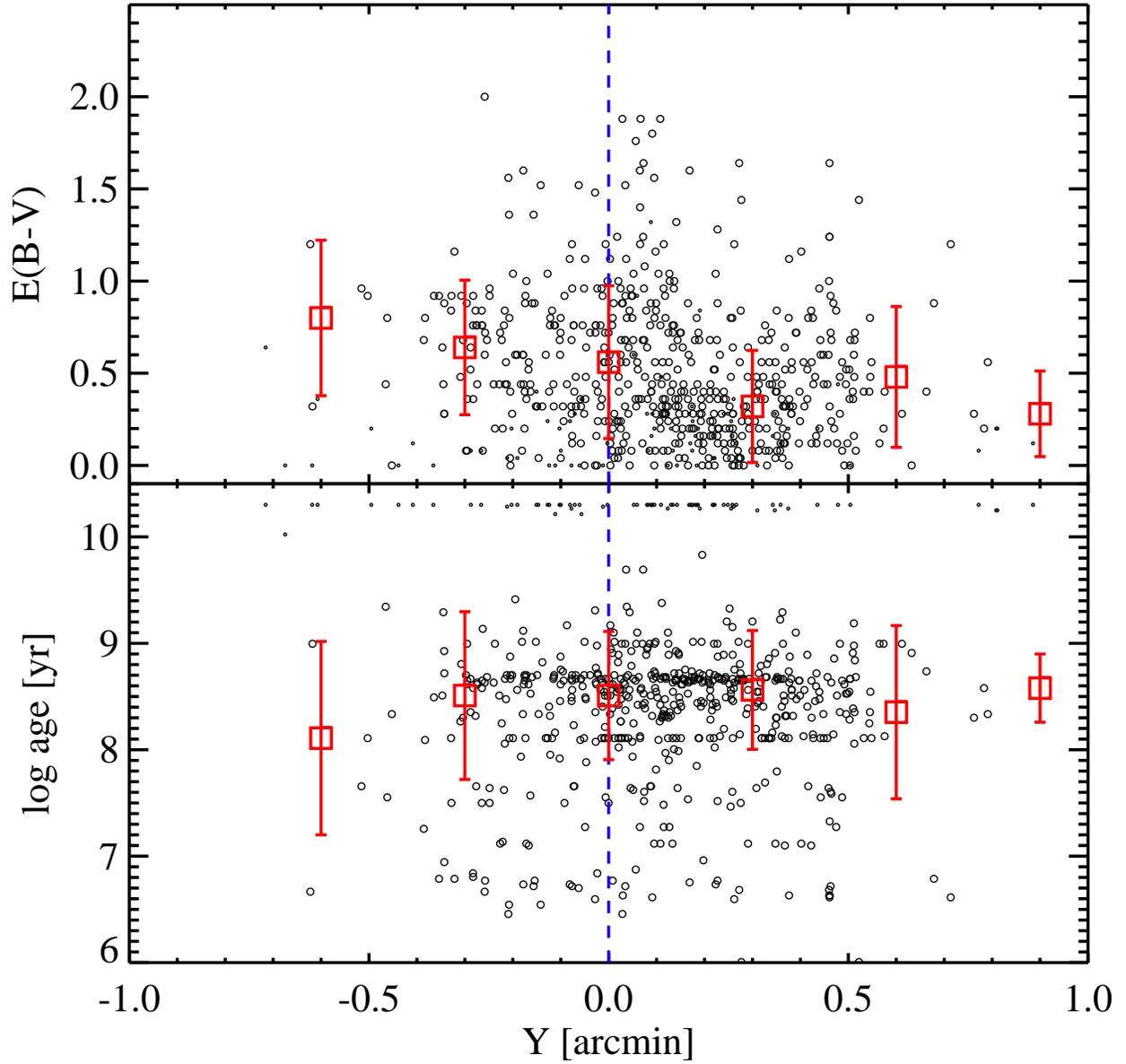} 
 \caption{ $E(B-V)$  (upper panel) and ages (lower panel) versus $Y$ for the M82 star clusters. The squares
with error bar represent the mean values of ages and $E(B-V)$ of the star clusters with log(age)$<10$.     
\label{ebvageY}}
\end{figure}

Figure \ref{ebvageX} and \ref{ebvageY} display variations of $E(B-V)$ and ages along the major axis $X$ and the minor axis $Y$ for the M82 star clusters.
$E(B-V)$ values are on average higher in the south-west side, indicating that the north-east side of the galaxy is farther than the south-west side. 
This result gives a hint for understanding the wider distribution of intermediate-color star clusters ($0.8\leqq (B-V)_0 <1.3$) in the north-east side. 
The star clusters with the largest values of $E(B-V)$ are located mostly in the nuclear region.
Mean values of $E(B-V)$ are 0.3--0.5 in the disk region, and $\sim 0.8$ in the nuclear region.
On the other hand, mean cluster ages show little variation along the $X$ or $Y$ axes.

We tested the effect of reddening errors to the derived ages and masses of the star clusters. The errors of derived $E(B- V)$ in our SED fitting for M82 clusters range from zero to about 0.1 with a mean value of 0.04. A reddening error of 0.1 for $E(B-V)$ leads to 0.1 dex difference in log(age) and negligible difference in mass.

\section{Discussion}

\subsection{Comparison with Previous Studies}

\begin{figure}
 \epsscale{1.0} \plotone{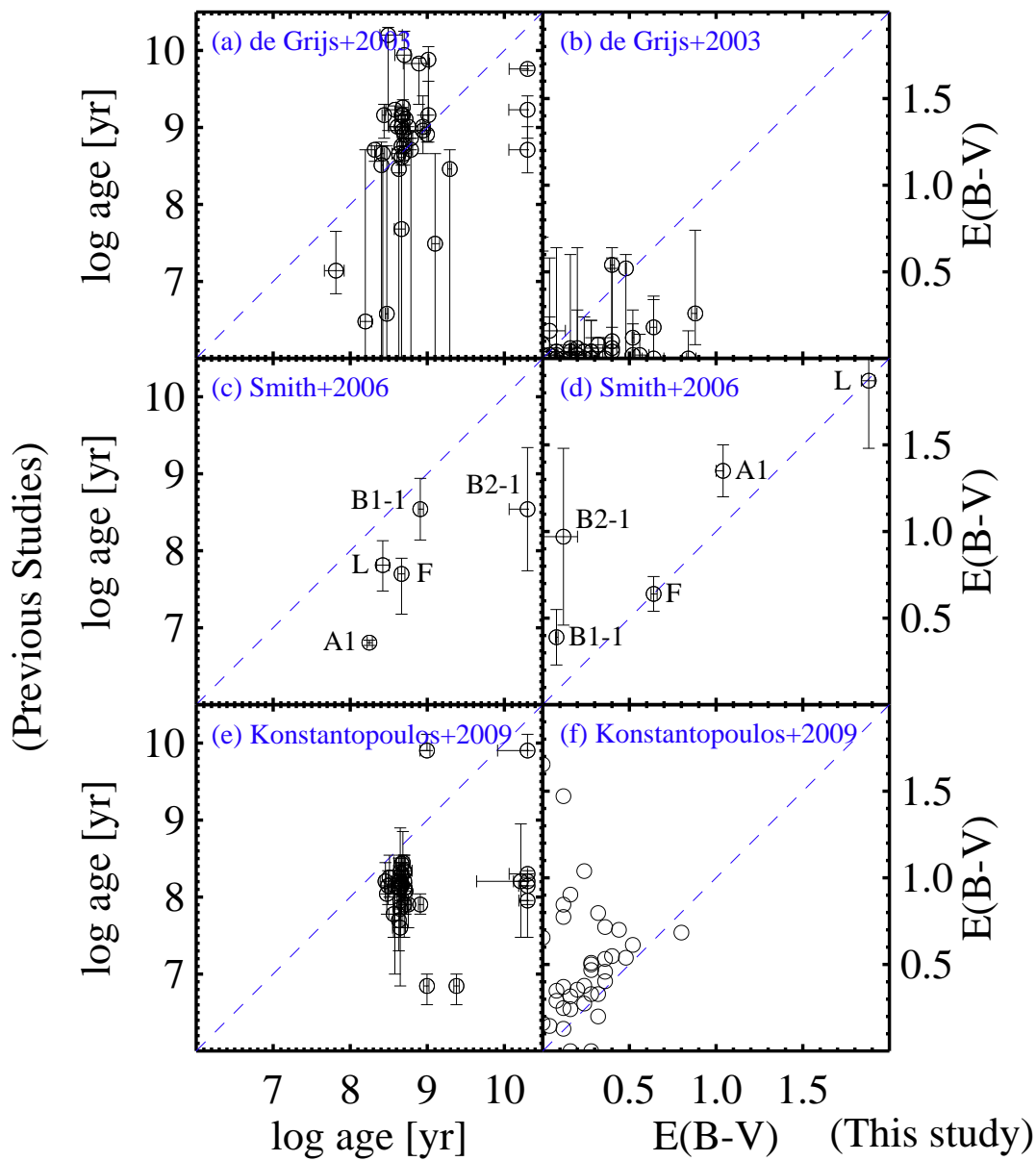} 
 \caption{ Comparison of ages (left panels) and $E(B-V)$ (right panels) derived in this study with previous studies  ((upper panels) \citet{deG03}, (middle panels) \citet{Smi06}, and (lower panels) \citet{Kon09}).
 Dashed lines represent one-to-one relations. 
\label{compage}}
\end{figure}

Ages and reddening values of star clusters in galaxies are often derived from the comparison of a combination of broad band photometry and/or spectra with SSP models.
In Figure \ref{compage} we compare our estimates for age and reddening with those in  \citet{deG03, Smi06, Kon09}.
Our ages and reddenings show a correlation with others's but with some scatters and systematic offsets. Our ages are
on average slightly larger ($\sim$0.7 dex) than \citet{Kon09}'s and \citet{Smi06}'s, but slightly smaller ($\sim$0.12 dex) than \citet{deG03}'s.
Our reddening values are
on average slightly smaller ($\sim$0.2) than \citet{Kon09}'s and \citet{Smi06}'s, but slightly larger ($\sim$0.2) than \citet{deG03}'s that are mostly zero values.
Note that \citet{Kon09} derived ages from the spectra first, then estimated reddenings from $BVI$ colors based on these ages.

These differences could be explained by raising a few points. 
The first possibility is the different filter sets. \citet{deG03} used $BVIJH$ filter set, while we used $UBVI(YJ)H$.
It is known that the existence of $U$ band data will lead to a better determination of ages and reddenings with SED fit methods, especially for young clusters. \citet{Smi07} found, from the $UBVI$ photometry of the star clusters in the region B, and that their ages are much younger than those of \citet{deG03}, while their reddenings are much larger than the latter. This is consistent with our findings as above. Note that \citet{deG03} obtained almost zero reddening values, which may lead to slightly older ages than those of ours ($\sim$0.2 dex) with estimated reddening of $E(B-V) \sim 0.3$. 

The second possibility is the different SSP models. We used SSP models from \citet{BC03}, while \citet{Smi06} and \citet{Kon09} adopted the STARBURST99 models from \citet{Lei99}. This model difference could lead to the different results of the age estimation.
Recently \citet{Rod11} investigated the evolution history of M82 using multi-wavelength photometry of unresolved stellar light in the disk.
They used 5 arcsec apertures to derive spatially resolved star
formation history of the M82 disk with multi-wavelength data from UV to NIR. They compared observational data with SED models from the Padova Group \citep{Bre93} with solar-metallicity and Salpeter IMF. This model is very similar to the model adopted in this study. They derived mean values for age and reddening of the apertures for the region B, ${\rm log (age)} = 8.60\pm0.06$ and $E(B-V)=0.44\pm0.05$. These values for integrated stellar light are similar to the mean values for the star clusters in the region B derived in this study, ${\rm log (age)} \approx 8.7$ and $E(B-V)=0.30$. Therefore the model difference could be a main source of age differences between our results and those of \citet{Smi06} and \citet{Kon09}.

The third possibility is the different methods. We used SED fitting with broad band photometry, while  \citet{Smi06} and \citet{Kon09} derived cluster parameters from spectroscopy. Both methods yield similar results in general, but with a large scatter. It is not easy to disentangle clearly each possibility among these.

\subsection{Halo Star Clusters in M82}

There is only a little information on the existence of old stellar population in the halo of M82, 
while there are many observational reports showing the existence of ionized gas, atomic gas, and molecular clouds in
the halo region of M82 \citep{Tay01,Chy08,Vei09,Rou10}.
Some observational studies reported the existence of intermediate-age stars outside the disk of M82 \citep{Dav08}.
Only a small number of globular clusters in the field of M82 were discovered by previous
spectroscopic studies (two in \citealp{Sai05} and two in \citealp{Kon09}). These globular clusters were found only in the south of M82 in the direction toward M81 so that \citet{May09} suspected
that they may not belong to M82.  

In this study, we have discovered 35 star clusters in the halo region of M82.
The color distribution of these halo clusters show a dominant blue peak at $(B-V) \approx 0.65$ and 
they are mostly old ($\gtrsim 10$ Gyr). In addition they are located in a wide area surrounding M82.
These suggest that a significant population of old globular clusters does exist in the halo of M82. 
However, there may be some contamination due to background galaxies.
Measurements of radial velocities of these objects are needed to
distinguish the halo members and the background galaxies.

\begin{figure}
 \epsscale{1.0} \plotone{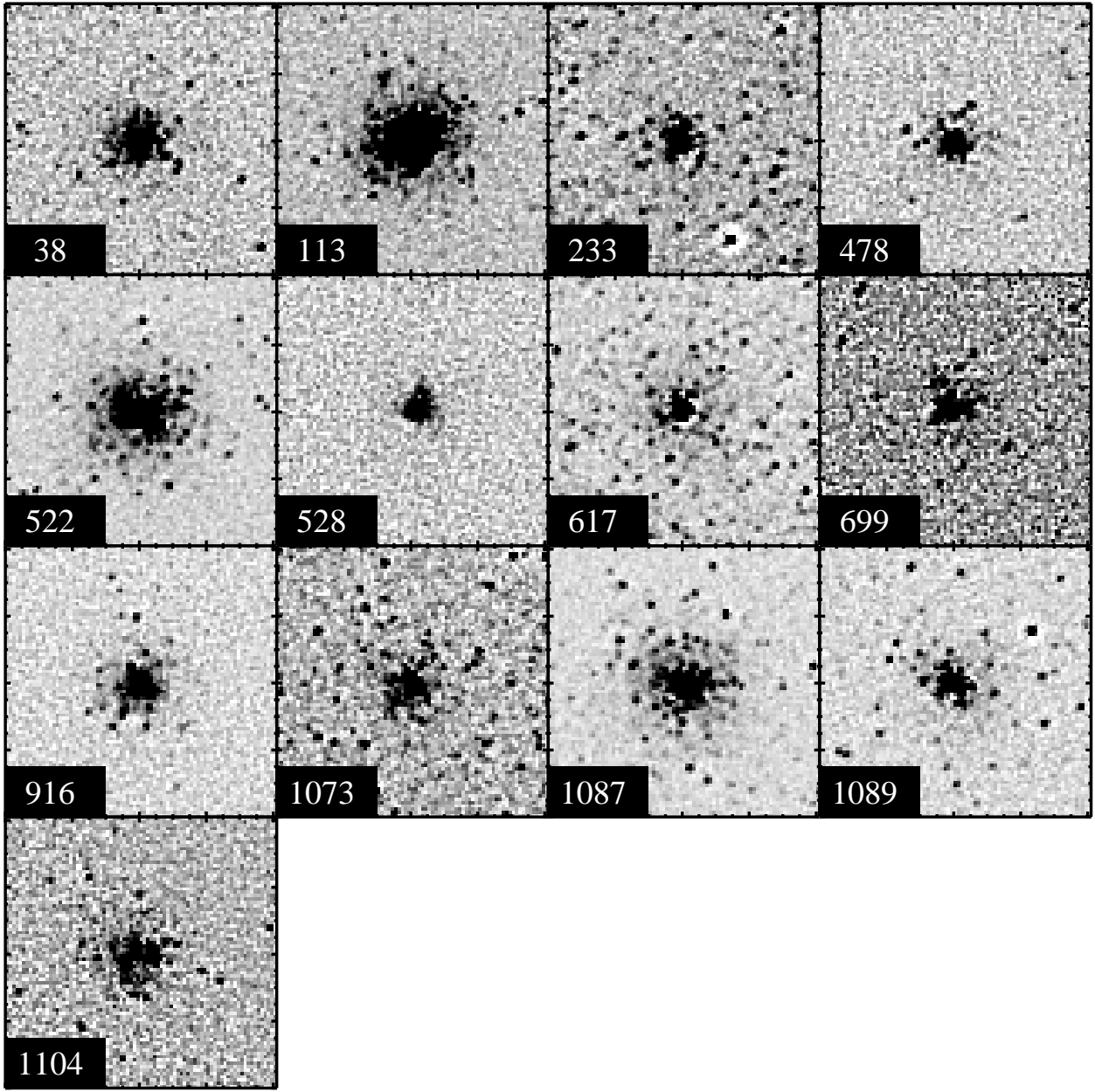} 
 \caption{Thumbnail $I$-band images of the halo star clusters  that are partially resolved into 
 individual stars. Each image covers $4\arcsec \times 4\arcsec$. 
 Smoothed images were subtracted from the original images 
 for  showing better 
 resolved stars. 
\label{rsl1}}
\end{figure}

Another noteworthy point for the halo clusters in M82 is that there are some clusters whose stars are partially resolved. 
With the resolving power of $HST$/ACS, it is possible to resolve some bright stars in M82 \citep{Lee13,jan12}. 
If we find star clusters that are partially resolved into stars in M82, then they can be considered to be the members of M82. 
We made a smoothed image for a field including each star cluster, and subtracted it from the original image to reveal any feature of resolved stars in each cluster.
Then we  inspected the subtracted images, finding 13 halo star clusters showing features of resolved stars, 
as shown in Figure\,\ref{rsl1}.

\begin{figure}
 \epsscale{1.0} \plotone{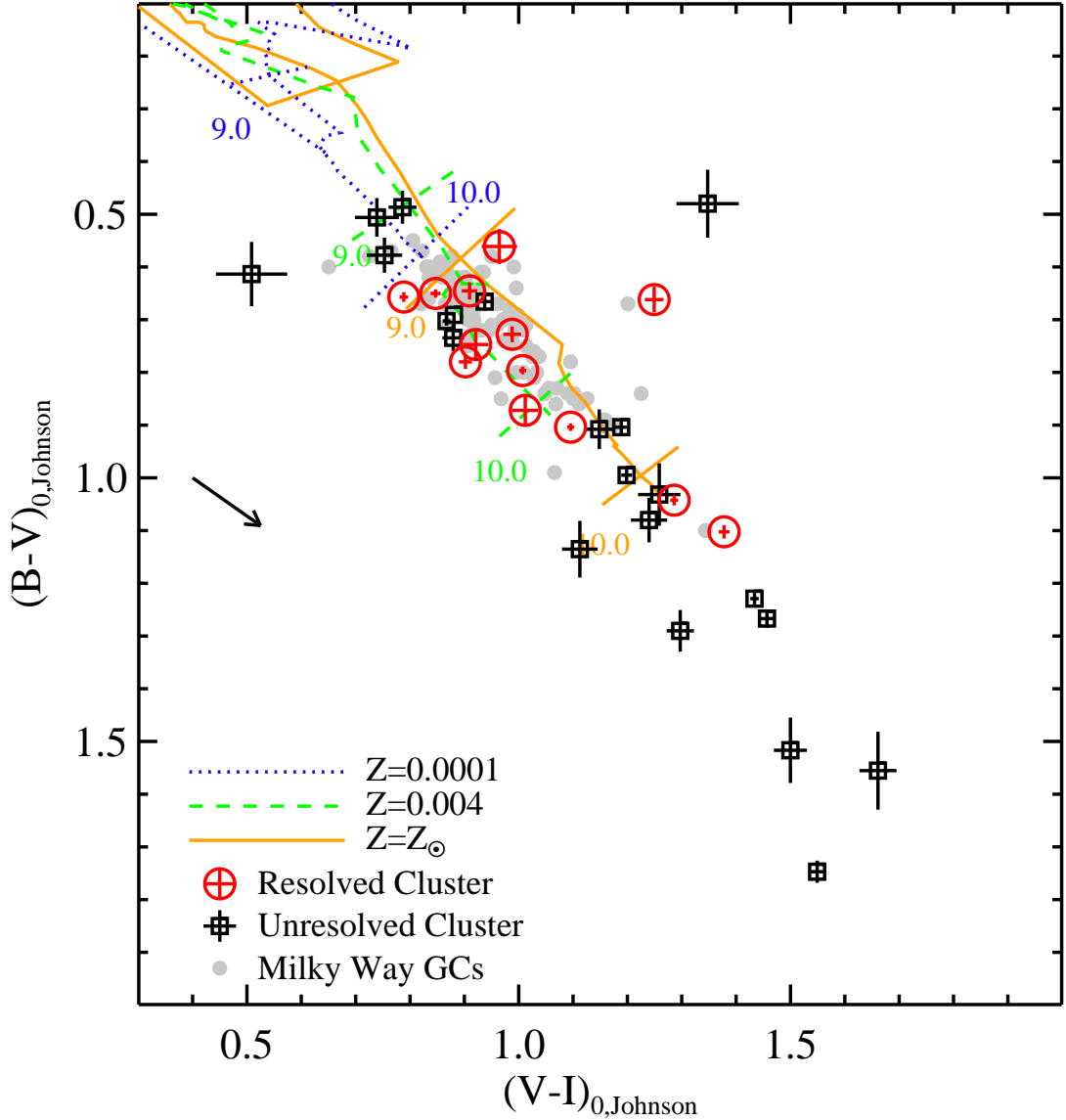} 
 \caption{$ (B-V)_{0, {\rm Johnson}}  -  (V-I)_{0, {\rm Johnson}} $ diagram of the star clusters in the halo region of M82. 
 Here $(B-V)_{0, {\rm Johnson}}$ and 
 $(V-I)_{0, {\rm Johnson}}$  are given in the Johnson-Cousins system, and foreground reddening is corrected.. 
 The size of 
 crosses represents an error of each color. Open circles and open squares indicate the 
 resolved and unresolved star clusters, respectively.
  Filled circles represent the globular clusters in the Milky Way Galaxy \citep{Har96}. 
  Dotted line, dashed line, and solid lines show the  
 simple stellar population models with $Z=0.0001$, 0.004, and 0.02, respectively, given in  \citep{Mar08}. 
Numbers labeled along  the model lines represent log(age).
 The arrow represents a reddening vector with $A_V=1$.
\label{ccd2}}
\end{figure}

We transformed the photometry of the halo star clusters from the $HST$ filter system into Johnson-Cousins filter system using the calibration information from \citet{Sir05} to compare the halo star clusters with the globular clusters in the Milky Way Galaxy \citep{Har96}. 
Figure\,\ref{ccd2} displays a $(B-V)_{0, {\rm Johnson}} - (V-I)_{0, {\rm Johnson}} $ CCD of the halo star clusters. SSP models for $Z=0.02, 0.004$ and 0.0001 \citep{BC03} are also plotted. 
The star clusters with and without resolved stars are
plotted in different symbols (circles and squares, respectively).
Most halo star clusters have color ranges with $0.5 \lesssim (B-V)_{0, {\rm Johnson}} \lesssim 1.2$ and
$0.8 \lesssim (V-I)_{0, {\rm Johnson}} \lesssim 1.5$.
This color range is similar to that of the Milky Way globular clusters. 
The resolved star clusters and unresolved star clusters have similar
color ranges except for a few extremely red unresolved clusters.
Therefore most of the unresolved star clusters are also probably the globular clusters of M82. 
Three reddest objects may be highly reddened star clusters. 

Since the halo region is far from the disk plane (projected distance $\gtrsim 880$ pc), we can assume that the internal extinction is negligible.
The halo population in late-type galaxies is usually metal poor so that we assume a lower metallicity for the halo star clusters in M82.
Based on these assumptions, the halo star clusters mostly concentrated
on the region at $(B-V)_0 \approx 0.65$ and $(V-I)_0 \approx 1.0$ in Figure 19 are 
consistent with 10 Gyr age with Z$=0.0001$. Several halo clusters are located in the region extended to the redder direction along the SSP model. These red halo clusters could be candidates of reddened old metal-rich star clusters.
Four of them are located relatively closer to  the disk ($\lesssim 1$kpc), so 
that they might have been reddened due to internal extinction.

\subsection{Comparison of M82 Halo Star Clusters with M81 Globular Clusters}

\begin{figure}
 \epsscale{1.0} \plotone{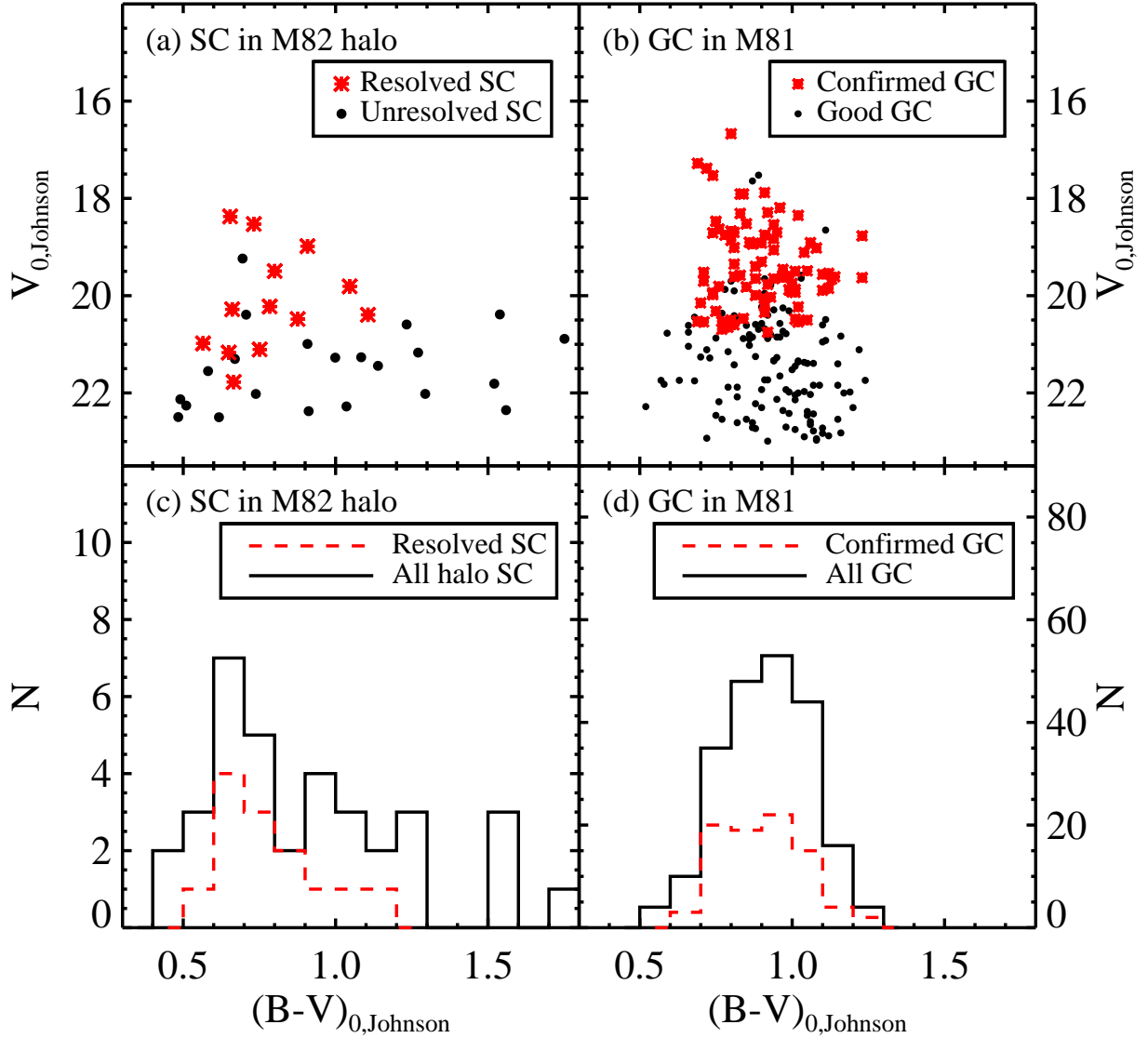} 
 \caption{ $ V_{0, {\rm Johnson}} - (B-V)_{0, {\rm Johnson}} $ diagrams (upper panels) and color histograms (lower panels) of the star clusters in the halo of M82 (left panels) in comparison with the globular clusters in M81 \citep{Nan11} (right panels). 
 Here $V_{0, {\rm Johnson}}$ and  $(V-I)_{0, {\rm Johnson}}$  are given in the Johnson-Cousins system, and foreground reddening is corrected. 
 Dots and asterisks in (a) represent unresolved and resolved 
 star clusters in the M82 halo, respectively. 
 Asterisks and dots in (b) indicate M81 globular clusters confirmed by spectroscopic studies, and
 good globular cluster candidates, respectively. 
  Solid and dashed 
 lines in (c) represent all star clusters and resolved star clusters in the M82 halo. 
  Solid and dashed lines in (d) represent all globular clusters and spectroscopically confirmed globular clusters  in M81.
\label{comp2} }
\end{figure}

We compared the M82 halo star clusters with the M81 globular clusters in the 
$V_{0, {\rm Johnson}} - (B-V)_{0, {\rm Johnson}}$ diagrams and $(B-V)_{0, {\rm Johnson}}$ color distributions in Figure\,\ref{comp2}.
It is relatively easy to select halo star clusters in M82,
because it is an almost edge-on galaxy. In contrast, it is difficult to select halo globular clusters in M81, because
M81 is almost face-on and many globular clusters were found in the projected plane of the disk region of M81. 

\citet{Nan11} found 214 globular cluster candidates in M81 including 85 confirmed by spectroscopic
studies \citep{Nan10a}, and  provided a catalog of the globular cluster candidates with Johnson-Cousins $BVI$ photometry.
The brightest globular clusters in M81 are about one magnitude brighter than those in M82.
The color ranges of the M82 halo star clusters and M81 globular clusters are similar, with
 $0.5 \lesssim (B-V)_{0, {\rm Johnson}} \lesssim 1.3$, although  four of the unresolved halo star clusters in M82 have redder colors than $(B-V)_0 \approx 1.3$.

The color distributions of the M82 and M81 clusters are different.
The M82 clusters show an asymmetric distribution, 
while the M81 clusters have a Gaussian-like distribution. 
The peak color for the M82 clusters,  $(B-V)_{0, {\rm Johnson}} \approx 0.65$, is bluer than that for the M81 clusters, 
at $(B-V)_{0, {\rm Johnson}} \approx 0.95$. This difference is seen both in the total sample and in the
confirmed globular cluster samples (resolved star clusters in M82 and spectroscopically confirmed globular 
clusters in M81).
If the internal extinction for the clusters is negligible and the mean ages are similar, the difference in the peak color 
could be explained by different metallicities:
the M82 halo clusters are on average more metal poor than
the M81 globular clusters. However it is noted that there are a small number of M82 halo clusters that have similar
colors to the peak color of the M81 globular clusters. 
These results show  that M82, showing a recent starburst, have an old stellar halo including metal poor old globular clusters, In addition, the origin of the metal-poor halo clusters in M82 appears to be different from that of the M81 globular clusters.

\subsection{Evolution of M82} 


M82 is a famous archetype of the starburst galaxy, and many studies have concentrated on understanding of the recent starburst.
Previous studies suggested that there might have been two starbursts in the central region of M82:
one at $\sim$10 Myr ago and another at $\sim$5 Myr ago \citep{For03}. The first starburst started 
in the nuclear region as well as in the bar, while the second burst in the circum-nuclear region.

Recently, \citet{May09} 
noted from the analysis of stellar light that M82 differs from normal late type galaxies: normal spiral galaxies mostly have disks  with continuing star formation, and its mean stellar age is old ($>5$ Gyr), while the mean age of the disk of M82 has relatively young ages ($\sim$500 Myr) and star formation was quenched at about 100 Myr ago. This is consistent with
the results derived from the star clusters in this study.
Young clusters in M82 have been used to study the impact driven by dynamical interactions with M81.
\citet{Cha01b} found 114 star clusters in eight WFPC2 fields of M81.
They showed that the young cluster formation in M81 started at $\sim$600 Myr ago from the analysis of $BVI$ CCD.
This is consistent with the population of star clusters with an age peak at $\sim$500 Myr in M82 derived in this study.
The existence of almost coeval star clusters in M81 and M82 suggests that the interaction
between two galaxies at this epoch may have enhanced the star cluster formation in both galaxies.

\citet{May09} suggested three possible scenarios 
to explain the star formation history derived from stellar light in the disk as follows:
(1) a starburst galaxy formed by flyby encounter of low surface brightness (LSB) galaxies, 
(2) a rejuvenated disk galaxy with  a truncated stellar disk, and
(3) a newly formed galaxy.
They supported the first scenario, but they could not rule out the other scenarios, either.
We discuss each scenario in the following considering the results derived in this study.

The first scenario is based on theoretical simulations \citep{Mih99}.
This simulation suggested that flyby encounter of a companion galaxy could have induced the disk-wide star formation in the host gas-rich LSB.
We found that a significant star formation over the entire disk at about 500 Myr ago from the age distribution of M82 disk clusters as shown in Figure \ref{age1}. 
If we adopt this model for the cause of the observed disk-wide star formation, then the epoch of the interaction between M81 and M82 should be about 500 Myr ago.
\citet{Yun93} suggested that the gas around M82 may have originated from M82, and it means that M82 was a gas-rich galaxy before its interaction with M81.
These results support this scenario. 
However, this scenario cannot explain the recent central starburst in M82. 

The second scenario was proposed by \citet{Sof98}.
In this scenario, the disk of M82 was stripped by the interaction and a new disk was formed. 
Then the old stars stripped from the disk and halo should be discovered in the M81 group.
These remnants of stripping were not yet found, but \citet{May09} noted that the stripped stars may be low-mass main sequence stars and they are fainter than the detection limit of $HST$ observation.
However, stellar ages and velocity fields of M82 are not consistent with the expectation of this scenario \citep{Kon09}. 
This scenario suggested that the stellar contents of M82 are dominated by the bulge population. However, we found that the star clusters with ages ranging from very young ($<10$ Myr) to very old ($>10$ Gyr) exist in the nuclear region. 
It suggests that the star clusters in the nuclear region are not dominated by the bulge population.

The last scenario is that M82 is a newly formed galaxy. 
However, we found the existence of the old globular cluster system in this study, and AGB stars are also found in the halo region \citep{Dav08}.  These show that M82 is not a newly formed galaxy. 
Therefore we can rule out the second and third scenarios.
Our results support the first scenario, but the existence of recent starbursts in the nuclear region
remains to be explained.


%

%
\section{Summary}
In this study we presented a photometric study of the star clusters in the starburst galaxy M82 based on high resolution $HST$ images. 
We found 1105 star clusters in the entire $HST$/ACS field of M82, and derived their photometry
from $UBVI(YJ)H$ images.
Primary results are summarized below.

\begin{itemize}

\item A majority of the star clusters are located in the disk of M82. The scale height of
their vertical number density profile is derived to be $h_z = 9\arcsec.64 \pm 0\arcsec.40$ ($164 \pm 7$ pc),
which is similar to that of the thin stellar disk of M82.
We also found 35 star clusters in the halo region of M82.

\item The CMDs and color distribution of the star clusters vary depending on the regions.
The star clusters in the disk of M82 are composed of a dominant blue population and a weaker red population.
The blue populations in the disk have a color peak at $(B-V)_0 \approx 0.45$, and the red populations in the disk are mostly redder than $(B-V)_0 \approx 1.0$.
The star clusters in the disk-edge region are mostly bluer than $(B-V)_0 \approx 0.8$, while those in the nuclear region have a large color range of $0.2 \lessapprox (B-V)_0 \lessapprox 2.0$. 
The star clusters in the halo region have a color range of $0.4 < (B-V)_0 \lessapprox 1.5$ that is similar to that of the globular clusters in the MWG.

\item The luminosity function of the disk star clusters follows a power-law function
with $\alpha=-2.04\pm0.03$ that is similar to those of young clusters in other galaxies, while the luminosity function of the star clusters in the halo and nuclear region do not follow power-law functions.
The luminosity function of the star clusters in the disk-edge region is also fit by a power-law function but with a
slightly steep slope, $\alpha=-2.77\pm0.21$.

\item The size distribution of the star clusters has a strong peak at $\sim2$ pc, and has a long tail to larger radii.
There is no significant trend between the star cluster size and galactic central distance.

\item We estimated the ages of about 630 star clusters in M82 with $UBVI(YJ)H$ band data, and found that the most dominant population shows a peak at $\sim500$ Myr. 
There are also some star clusters with younger ($<10$ Myr) and older ($>10$ Gyr) ages.

\item The mass function of the young star clusters (log(age)$<7.0$) is fit by a power law with a slope ($\beta\sim-1.53\pm0.08$) shallower compared with that of other late type galaxies such as M51 and Antennae galaxies. 

\item We found two populations of star clusters in the nuclear region: very young clusters with ages of 5 Myr and intermediate age clusters with ages of $\sim$500 Myr. The former represents
the recent starburst, while the latter are consistent with those in the disk.
The brightest young clusters have high masses ($>10^5 $M$_{\odot}$), indicating that massive clusters form preferentially in the high density regions.  

\item The colors and ages of the halo star clusters in M82 are similar to those of the globular clusters in the Milky Way Galaxy, so that they could be regarded as old globular clusters in M82. 
It suggests that M82 has an old stellar halo that is similar to that of other normal galaxies.
From the comparison of $(B-V)_{0, {\rm Johnson}}$ colors, we have found that the M82 globular cluster system is on average more metal poor than M81 globular cluster system.

\item Our results suggest that there were four epochs of starburst in M82. The first epoch is when the old globular clusters were formed.
The second is when a majority of the star clusters were formed in the entire disk, which started 1 Gyr ago and stopped 100 Myr ago. 
The third is when star clusters were formed in the nuclear region $<10$ Myr ago.
The fourth is when star clusters were formed recently in the central region (maybe $\sim$3 Myr ago).

\end{itemize}

\acknowledgments
This work was supported by the National Research Foundation of Korea (NRF) grant
funded by the Korea Government (MEST) (No. 2012R1A4A1028713).




\clearpage

\makeatletter
\setlength{\abovecaptionskip}{6pt}   
\setlength{\belowcaptionskip}{6pt}   
\long\def\@makecaption#1#2{%
  \vskip\abovecaptionskip
  \sbox\@tempboxa{#1 #2}%
  \ifdim \wd\@tempboxa >\hsize
    #1 #2\par
  \else
    \global \@minipagefalse
    \hb@xt@\hsize{\box\@tempboxa\hfil}%
  \fi
  \vskip\belowcaptionskip}
\makeatother

\clearpage







\begin{table}
\begin{center}
\caption{A summary of data.\label{tbl-1}}
\begin{tabular}{lccc}
\tableline\tableline Filter & Proposal ID & Instrument & Exposure times
(s)
\\
\tableline
$F435W (B)$ & 10776 & $HST$/ACS WFC & 1600 \\
$F555W (V)$ & 10776 & $HST$/ACS WFC & 1360 \\
$F814W (I)$ & 10776 & $HST$/ACS WFC & 1360 \\
\tableline
$F336W (U)$ & 11360 & $HST$/WFC3 UVIS & 1050, 1215, 1620 \\
$F110W (YJ)$ & 11360 & $HST$/WFC3 IR & 598, 1195 \\ 
$F160W (H)$ & 11360 & $HST$/WFC3 IR & 598, 2395 \\ 
\tableline
\end{tabular}
\end{center}
\end{table}

\clearpage

\begin{landscape}
\begin{table}
\tiny
\begin{center}
\caption{A catalog of the star clusters in M82 \label{tbl-2}}
\begin{tabular}{lllcccccccc}
\tableline\tableline ID & R.A.(J2000) & Dec.(J2000) & $V$ & ($B-V$) &
 ($V-I$) & ($U-B$) & ($I-YJ$) & ($YJ-H$) & $r_{eff}$ [pc] & log(age) [yr] \\
\tableline
 481 &  9 55 47.07 &  69 40 42.3 & 16.289 $\pm$  0.007 &  1.127 $\pm$  0.012 &  1.531 $\pm$  0.009 &  0.764 $\pm$  0.015 &  1.050 $\pm$  0.013 &  0.697 $\pm$  0.019 &  ...  &  8.67 \\
  611 &  9 55 52.68 &  69 40 46.0 & 17.061 $\pm$  0.028 &  1.032 $\pm$  0.044 &  0.906 $\pm$  0.092 & -0.139 $\pm$  0.038 &  0.545 $\pm$  0.353 &  1.140 $\pm$  0.368 &  ...  &  9.17 \\
  628 &  9 55 53.47 &  69 40 51.1 & 17.125 $\pm$  0.004 &  1.322 $\pm$  0.006 &  2.061 $\pm$  0.013 &  0.500 $\pm$  0.009 &  1.166 $\pm$  0.051 &  0.815 $\pm$  0.069 &  3.23 &  8.25 \\
  624 &  9 55 53.34 &  69 40 49.8 & 17.207 $\pm$  0.020 &  1.128 $\pm$  0.021 &  1.425 $\pm$  0.075 &  0.382 $\pm$  0.011 &  1.162 $\pm$  0.194 &  0.738 $\pm$  0.260 &  4.76 &  9.69 \\
  561 &  9 55 50.42 &  69 40 47.3 & 17.407 $\pm$  0.013 &  1.260 $\pm$  0.017 &  1.521 $\pm$  0.047 &  0.366 $\pm$  0.016 &  1.267 $\pm$  0.098 &  0.618 $\pm$  0.139 &  4.91 &  9.38 \\

\tableline
\tablenotetext{}{(This table is available in its entirety in a machine-readable form in the online journal. A portion is shown here for guidance regarding its form and content.)}
\end{tabular}
\end{center}
\end{table}
\end{landscape}

\clearpage


\clearpage

\clearpage

\clearpage

\clearpage

\clearpage

\clearpage

\clearpage

\clearpage

\clearpage

\clearpage

\clearpage

\begin{thebibliography}{}

\bibitem[Anders
\& Fritze-v.~Alvensleben(2003)]{And03} Anders, P., \& Fritze-v.~Alvensleben, U.\ 2003, \aap, 401, 1063

\bibitem[Anders et al.(2004)]{And04} Anders, P., Bissantz, N., Fritze-v.~Alvensleben,
U., \& de Grijs, R.\ 2004, \mnras, 347, 196

\bibitem[Annibali et al.(2011)]{Ann11} Annibali, F., Tosi, 
M., Aloisi, A., \& van der Marel, R.~P.\ 2011, \aj, 142, 129 

%
\bibitem[Barker, de Grijs,
\& Cervi{\~n}o(2008)]{Bar08} Barker, S., de Grijs, R., \& Cervi{\~n}o, M.\
2008, \aap, 484, 711
%
%
\bibitem[Bastian, Covey, \& Meyer (2010)]{bas10} Bastian, N., Covey, K. R., \& Meyer, M. R. 2010, \araa, 48, 339 
\bibitem[Bastian et al.(2012)]{Bas12} Bastian, N., Adamo, A.,
Gieles, M., et al.\ 2012, \mnras, 419, 2606 


\bibitem[Bell
\& Kennicutt(2001)]{Bel01} Bell, E.~F., \& Kennicutt, R.~C., Jr.\ 2001, \apj, 548, 681


\bibitem[Bertin
\& Arnouts(1996)]{Ber96} Bertin, E., \& Arnouts, S.\ 1996, \aaps, 117, 393
%
%
\bibitem[Bertin et al.(2002)]{Ber02} Bertin, E., Mellier, Y.,
Radovich, M., Missonnier, G., Didelon, P.,
\& Morin, B.\ 2002, Astronomical Data Analysis Software and Systems XI,
281, 228
%
%
\bibitem[Bressan et
al.(1993)]{Bre93} Bressan, A., Fagotto, F., Bertelli, G., \& Chiosi, C.\ 1993, \aaps, 100, 647

\bibitem[Bruzual
\& Charlot(2003)]{BC03} Bruzual, G., \& Charlot, S.\ 2003, \mnras, 344,
1000
%
%
\bibitem[Cantiello et
al.(2009)]{Can09} Cantiello, M., Brocato, E., \& Blakeslee, J.~P.\ 2009, \aap, 503, 87

\bibitem[Chabrier(2003)]{Cha03} Chabrier, G.\ 2003, \pasp, 
115, 763 


\bibitem[Chandar et al.(2010)]{Cha10} Chandar, R., Whitmore,
B.~C., Kim, H., et al.\ 2010, \apj, 719, 966


%
%
\bibitem[Chandar, Tsvetanov,
\& Ford(2001)]{Cha01b} Chandar, R., Tsvetanov, Z., \& Ford, H.~C.\ 2001,
\aj, 122, 1342
%
\bibitem[Chandar et al.(2004)]{Cha04} Chandar, R., Whitmore,
B., \& Lee, M.~G.\ 2004, \apj, 611, 220
%
\bibitem[Chynoweth et al.(2008)]{Chy08} Chynoweth, K.~M., Langston,
G.~I., Yun, M.~S., Lockman, F.~J., Rubin, K.~H.~R.,
\& Scoles, S.~A.\ 2008, \aj, 135, 1983
%
%
\bibitem[Davidge(2008)]{Dav08} Davidge, T.~J.\ 2008, \aj, 136, 2502
%
%
\bibitem[de Grijs, Bastian,
\& Lamers(2003)]{deG03} de Grijs, R., Bastian, N., \& Lamers,
H.~J.~G.~L.~M.\ 2003, \mnras, 340, 197
%
%
\bibitem[de Grijs, O'Connell,
\& Gallagher(2001)]{deG01} de Grijs, R., O'Connell, R.~W., \& Gallagher,
J.~S., III 2001, \aj, 121, 768
%
\bibitem[de Grijs 
\& van der Kruit(1996)]{deG96} de Grijs, R., \& van der Kruit, P.~C.\ 1996, \aaps, 117, 19 

%
\bibitem[de Vaucouleurs et al.(1991)]{deV91} de Vaucouleurs,
G., de Vaucouleurs, A., Corwin, H.~G., Jr., Buta, R.~J., Paturel, G.,
\& Fouque, P.\ 1991, Volume 1-3, XII, 2069 pp.~7 figs..~ Springer-Verlag
Berlin Heidelberg New York,
%
\bibitem[Durrell, Sarajedini, \& Rupali (2010)]{dur10} Durrell, P.~R., Sarajedini, A., \& Chandar, R.\ 2010, \apj, 718, 1118  

%
%
%
%
\bibitem[F{\"o}rster Schreiber, Genzel, Lutz,
\& Sternberg(2003)]{For03} F{\"o}rster Schreiber, N.~M., Genzel, R., Lutz,
D., \& Sternberg, A.\ 2003, \apj, 599, 193
%
%
%
%
%
%
%
%
%
\bibitem[Gerke et al. (2011)]{ger11}Gerke, J. R., Kochanek, C. S., Prieto, J. L., Stanek, K. Z., \& Macri, L. M. 2011, arXiv:1103.0549  

\bibitem[Girardi et
al.(1995)]{Gir95} Girardi, L., Chiosi, C., Bertelli, G., \& Bressan, A.\ 1995, \aap, 298, 87

\bibitem[Gonz{\'a}lez Delgado et al.(2005)]{Gon05}
Gonz{\'a}lez Delgado, R.~M., Cervi{\~n}o, M., Martins, L.~P., Leitherer,
C., \& Hauschildt, P.~H.\ 2005, \mnras, 357, 945

\bibitem[Greco, Martini, \& Thompson (2012)]{gre12}Greco, J. P.,   Martini, P., \& Thompson, T. A. 2012, \apj, 757, 24

\bibitem[Harris(1996)]{Har96} Harris, W.~E.\ 1996, \aj, 112,
1487

\bibitem[Hwang
\& Lee(2008)]{Hwa08} Hwang, N., \& Lee, M.~G.\ 2008, \aj, 135, 1567

\bibitem[Hwang
\& Lee(2010)]{Hwa10} Hwang, N., \& Lee, M.~G.\ 2010, \apj, 709, 411
%
%
\bibitem[Ichikawa et al.(1995)]{Ich95} Ichikawa, T., Yanagisawa, K.,
Itoh, N., Tarusawa, K., van Driel, W., \& Ueno, M.\ 1995, \aj, 109, 2038
%
\bibitem[Jang et al.(2012)]{jan12}
Jang, I. S., Lim,. S., Park, H. S., \& Lee, M. G. 2012, 
\apj, 751, L19
%
\bibitem[Konstantopoulos et al.(2008)]{Kon08}
Konstantopoulos, I.~S., Bastian, N., Smith, L.~J., et al.\ 2008, \apj, 674,
846

\bibitem[Konstantopoulos et al.(2009)]{Kon09}
Konstantopoulos, I.~S., Bastian, N., Smith, L.~J., Westmoquette, M.~S.,
Trancho, G., \& Gallagher, J.~S. III\ 2009, \apj, 701, 1015
%
%
\bibitem[Lan{\c c}on et al.(2008)]{Lan08} Lan{\c c}on, A., Gallagher,
J.~S., III, Mouhcine, M., Smith, L.~J., Ladjal, D.,
\& de Grijs, R.\ 2008, \aap, 486, 165
%
%
\bibitem[Larsen(1999)]{Lar99} Larsen, S.~S.\ 1999, \aaps, 139, 393

\bibitem[Larsen(2002)]{Lar02} Larsen, S.~S.\ 2002, \aj, 124,
1393

\bibitem[Larsen et
al.(2011)]{Lar11} Larsen, S.~S., de Mink, S.~E., Eldridge, J.~J., et al.\ 2011, \aap, 532, A147

\bibitem[Lee et al.(2013)]{Lee13} Lee, M. G. et al. 2013, in preparation
%
%
%
\bibitem[Leitherer et al.(1999)]{Lei99} Leitherer, C.,
Schaerer, D., Goldader, J.~D., et al.\ 1999, \apjs, 123, 3


\bibitem[Lynds \& Sandage(1963)]{Lyn63} Lynds, C.~R., \& Sandage, A.~R.\ 1963, \aj, 68,
284
%
\bibitem[Marigo et
al.(2008)]{Mar08} Marigo, P., Girardi, L., Bressan, A., et al.\ 2008, \aap, 482, 883
%
\bibitem[Ma{\'{\i}}z Apell{\'a}niz(2009)]{Mai09} Ma{\'{\i}}z
Apell{\'a}niz, J.\ 2009, \apj, 699, 1938
%

%

%
\bibitem[Mayya, Carrasco, \& Luna(2005)]{May05} Mayya, Y.~D., Carrasco, L., \& Luna, A.\ 2005,
\apjl, 628, L33
%
\bibitem[Mayya et al. (2006)]{may06} Mayya, Y. D., Bressan, A., Carrasco, L., \& Hernandez-
Martinez, L. 2006, \apj, 649, 172   

\bibitem[Mayya et al.(2008)]{May08} Mayya, Y.~D., Romano, R.,
Rodr{\'{\i}}guez-Merino, L.~H., Luna, A., Carrasco, L.,
\& Rosa-Gonz{\'a}lez, D.\ 2008, \apj, 679, 404
%
\bibitem[Mayya \& Carrasco(2009)]{May09} Mayya, Y.~D., \& Carrasco, L.\ 2009, Revista
Mexicana de Astronomia y Astrofisica Conference Series, 37, 44
%
%
\bibitem[Melo et al.(2005)]{Mel05} Melo, V.~P.,
Mu{\~n}oz-Tu{\~n}{\'o}n, C., Ma{\'{\i}}z-Apell{\'a}niz, J.,
\& Tenorio-Tagle, G.\ 2005, \apj, 619, 270
%
\bibitem[Mihos(1999)]{Mih99} Mihos, J.~C.\ 1999, IAU Symp 186 Galaxy
Interactions at Low and High Redshift, 205
%
\bibitem[Mutchler et al.(2007)]{Mut07} Mutchler, M., et al.\ 2007,
\pasp, 119, 1
%
%
\bibitem[Nantais
\& Huchra(2010)]{Nan10a} Nantais, J.~B., \& Huchra, J.~P.\ 2010, \aj, 139,
2620
%
%
%
%
\bibitem[Nantais et al.(2011)]{Nan11} Nantais, J.~B., Huchra,
J.~P., Zezas, A., Gazeas, K., \& Strader, J.\ 2011, \aj, 142, 183

\bibitem[O'Connell
\& Mangano(1978)]{OC78} O'Connell, R.~W., \& Mangano, J.~J.\ 1978, \apj,
221, 62
%
%
\bibitem[O'Connell et al.(1995)]{OC95} O'Connell, R.~W., Gallagher, J.~S., III, Hunter,
D.~A., \& Colley, W.~N.\ 1995, \apjl, 446, L1
%
\bibitem[Ohyama et al. (2002)]{oya02} Ohyama et al. 2002, \pasj, 54, 891
%
\bibitem[Park et al.(2009)]{Par09} Park, W.-K., Park, H.~S.,
\& Lee, M.~G.\ 2009, \apj, 700, 103

\bibitem[Pellerin et al.(2010)]{Pel10} Pellerin, A., Meurer,
G.~R., Bekki, K., et al.\ 2010, \aj, 139, 1369
\bibitem[Rieke et al. (1980)]{rie80}Rieke, G. H., Lebofsky, M. J., Thompson, R. I., Low, F.
J., \& Tokunaga, A. T. 1980, \apj, 238, 24

\bibitem[Rodr{\'{\i}}guez-Merino et al.(2011)]{Rod11}
Rodr{\'{\i}}guez-Merino, L.~H., Rosa-Gonz{\'a}lez, D.,
\& Mayya, Y.~D.\ 2011, \apj, 726, 51

\bibitem[Roussel et al.(2010)]{Rou10} Roussel, H., et al.\ 2010,
\aap, 518, L66
%
%
\bibitem[Saito et al.(2005)]{Sai05} Saito, Y., et al.\ 2005, \apj, 621,
750
%
\bibitem[Salpeter(1955)]{Sal55} Salpeter, E.~E.\ 1955, \apj,
121, 161
%
\bibitem[Santiago-Cort{\'e}s, Mayya,
\& Rosa-Gonz{\'a}lez(2010)]{San10} Santiago-Cort{\'e}s, M., Mayya, Y.~D.,
\& Rosa-Gonz{\'a}lez, D.\ 2010, \mnras, 405, 1293
%
%
\bibitem[Scheepmaker et al.(2007)]{Sch07} Scheepmaker, R.~A.,
Haas, M.~R., Gieles, M., Bastian, N., Larsen, S.~S.,
\& Lamers, H.~J.~G.~L.~M.\ 2007, \aap, 469, 925
%
%
\bibitem[Schlegel, Finkbeiner,
\& Davis(1998)]{Sch98} Schlegel, D.~J., Finkbeiner, D.~P., \& Davis, M.\
1998, \apj, 500, 525
%
\bibitem[Shen \& Lo (1995)]{she95}Shen, J., \& Lo, K. Y. 1995, ApJ, 445, L99
%
\bibitem[Sirianni et al.(2005)]{Sir05} Sirianni, M., et al.\ 2005,
\pasp, 117, 1049
%

%
%
\bibitem[Smith et al.(2006)]{Smi06} Smith, L.~J., Westmoquette, M.~S.,
Gallagher, J.~S., O'Connell, R.~W., Rosario, D.~J.,
\& de Grijs, R.\ 2006, \mnras, 370, 513
%
%
\bibitem[Smith et al.(2007)]{Smi07} Smith, L.~J., et al.\ 2007, \apjl,
667, L145
%
\bibitem[Sofue(1998)]{Sof98} Sofue, Y.\ 1998, \pasj, 50, 227
%
%
\bibitem[Taylor, Walter,
\& Yun(2001)]{Tay01} Taylor, C.~L., Walter, F., \& Yun, M.~S.\ 2001,
\apjl,
562, L43
%
\bibitem[Telesco et al. (1991)]{tel91}Telesco, C. M., Joy, M., Dietz, K., Decher, R., \& Campins, H. 1991, \apj, 369, 135 

\bibitem[Vansevi{\v c}ius et al.(2009)]{Van09} Vansevi{\v
c}ius, V., Kodaira, K., Narbutis, D., et al.\ 2009, \apj, 703, 1872
%
\bibitem[Veilleux, Rupke,
\& Swaters(2009)]{Vei09} Veilleux, S., Rupke, D.~S.~N., \& Swaters, R.\
2009, \apjl, 700, L149
%
%
\bibitem[Whitmore et al.(2010)]{Whi10} Whitmore, B.~C., et al.\
2010, \aj, 140, 75
%
%
\bibitem[Whitmore et al.(1999)]{Whi99} Whitmore, B.~C., Zhang, Q.,
Leitherer, C., Fall, S.~M., Schweizer, F.,
\& Miller, B.~W.\ 1999, \aj, 118, 1551
%
%
\bibitem[Wills et al.(2000)]{Wil00} Wills, K.~A., Das, M.,
Pedlar, A., Muxlow, T.~W.~B., \& Robinson, T.~G.\ 2000, \mnras, 316, 33
%
%
%
%
\bibitem[Yun(1999)]{Yun99} Yun, M.~S.\ 1999, IAU Symp 186 Galaxy Interactions at Low
and High Redshift, 81
%
\bibitem[Yun et al.(1993)]{Yun93} Yun, M.~S., Ho, P.~T.~P.,
\& Lo, K.~Y.\ 1993, \apjl, 411, L17

%
\bibitem[Yun, Ho,
\& Lo(1994)]{Yun94} Yun, M.~S., Ho, P.~T.~P., \& Lo, K.~Y.\ 1994, \nat,
372, 530
%
%


\end{thebibliography}
\end{document}